\documentclass[sigconf]{acmart}

\usepackage{multirow}
\usepackage[utf8]{inputenc}
\usepackage{textcomp}

\usepackage{framed}
\usepackage{listings}

\AtBeginDocument{%
  }

\copyrightyear{2026}
\acmYear{2026}
\setcopyright{cc}
\setcctype{by}
\acmConference[CHI '26]{Proceedings of the 2026 CHI Conference on Human Factors in Computing Systems}{April 13--17, 2026}{Barcelona, Spain}
\acmBooktitle{Proceedings of the 2026 CHI Conference on Human Factors in Computing Systems (CHI '26), April 13--17, 2026, Barcelona, Spain}
\acmPrice{}
\acmDOI{10.1145/3772318.3791391}
  
\acmISBN{979-8-4007-2278-3/2026/04}

\begin{document}

\title{CritiqueCrew: Orchestrating Multi-Perspective Conversational Design Critique}

\author{Xiaojiao Chen}
\orcid{0000-0002-9276-1063}
\affiliation{%
  \institution{Laboratory of Art and Archaeology Image, Ministry of Education, Zhejiang University}
  \city{Hangzhou}
  \state{Zhejiang}
  \country{China}
}
\email{chenxiaojiao@zju.edu.cn}

\author{Jiahuan Zhou}
\affiliation{%
  \institution{Laboratory of Art and Archaeology Image, Ministry of Education, Zhejiang University}
  \city{Hangzhou}
  \state{Zhejiang}
  \country{China}
 }
\email{12021203@zju.edu.cn}

\author{Yunfeng Shu}
\authornote{Corresponding author.}
\orcid{0000-0002-1380-6586}
\affiliation{
\department{Laboratory of Art and Archaeology Image, Ministry of Education}\institution{Zhejiang University}
\city{Hangzhou}
\state{Zhejiang}
\country{China}}
\email{yfshu@zju.edu.cn}


\author{Ruihan Wang}
\affiliation{%
  \institution{Laboratory of Art and Archaeology Image, Ministry of Education, Zhejiang University}
  \city{Hangzhou}
  \state{Zhejiang}
  \country{China}
 }
\email{ruihanwang@zju.edu.cn}

\author{Qinghua Liu}
\affiliation{%
  \institution{Laboratory of Art and Archaeology Image, Ministry of Education, Zhejiang University}
  \city{Hangzhou}
  \state{Zhejiang}
  \country{China}
 }
\email{nico_yazawa@zju.edu.cn}

\begin{teaserfigure}
  \centering
  \includegraphics[width=\textwidth]{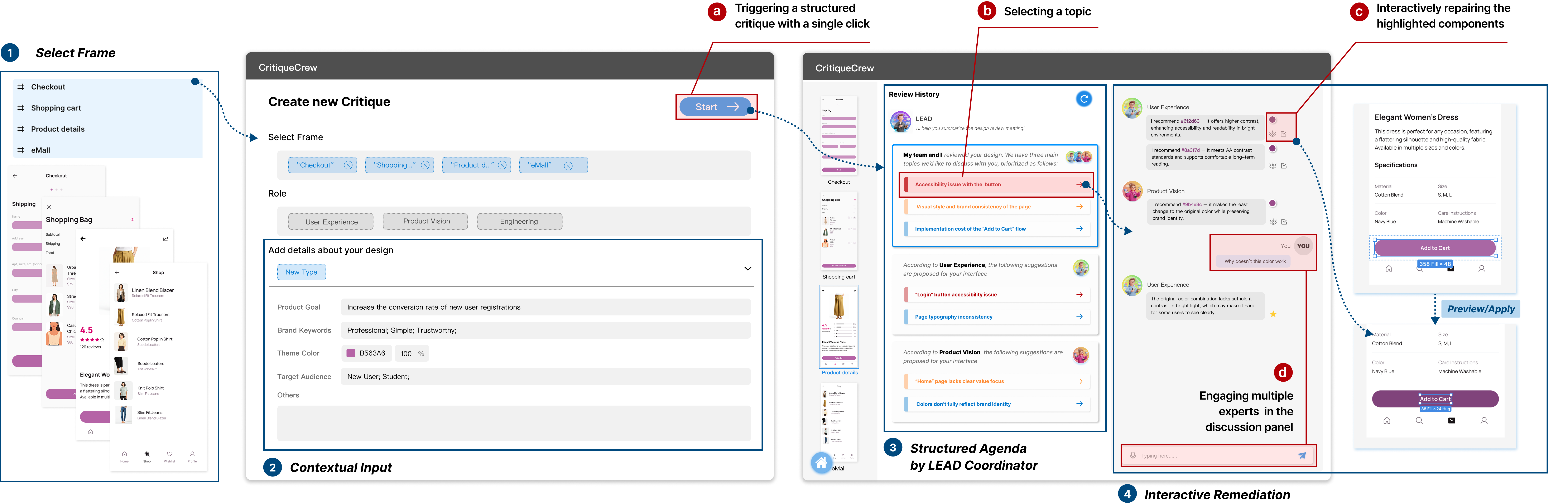}
  \caption{CritiqueCrew embeds a multi-perspective critique process directly in Figma. Designers first select target frames \includegraphics[height=1.2em]{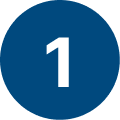} and provide contextual inputs such as product goals and brand keywords \includegraphics[height=1.2em]{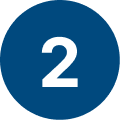}. Critiques from multiple expert roles are then synthesized into a structured agenda by the Lead Coordinator \includegraphics[height=1.2em]{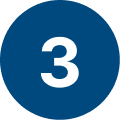}. Through in-context remediation tools and multi-perspective dialogue, designers can seamlessly explore trade-offs and apply modifications \includegraphics[height=1.2em]{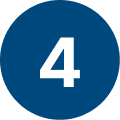}. This pipeline transforms critique from static checklists into an ongoing, explainable collaboration that empowers designers with actionable insights.}
  \Description{The figure illustrates the four-step critique workflow of CritiqueCrew within Figma. Step 1 (Select Frame): The designer chooses a target design frame (e.g., a checkout screen) from the Figma canvas. Step 2 (Contextual Input): A panel allows the user to input design goals, brand keywords, and target audience to guide the critique. Step 3 (Structured Agenda): The Lead Coordinator synthesizes feedback from multiple Expert Roles (user experience (UX), product vision, engineering) into a prioritized list of issues, displayed in a sidebar. Step 4 (Interactive Remediation): The user can select an issue to see detailed comments and engage in dialogue. The system visually highlights the relevant UI component on the canvas and offers a one-click "Apply" button to implement suggested fixes. Overall, the figure shows how the system transforms static critique into an actionable, multi-perspective workflow.}
\end{teaserfigure}

\renewcommand{\shortauthors}{Chen et al.}

\begin{abstract}
UI designers face growing cognitive load and cross-functional friction at the intersection of user needs, business goals, and engineering constraints. Existing automated tools often deliver static ``problem lists,'' lacking actionable repair paths and disrupting creative flow. We introduce CritiqueCrew, a Figma tool that supports designers through conversational critique. CritiqueCrew generates multi-faceted insights by implementing a multi-perspective orchestration of distinct expert roles (UX, PM, Engineer). It translates abstract critiques into concrete actions via in-context feedback and interactive remediation. Across two independent controlled studies (Total $N=48$), CritiqueCrew significantly improved both design quality and subjective experience compared to a traditional static checker. Furthermore, our results confirm that the structured orchestration of expert roles—rather than a unified model—is key to fostering trust and creativity support. Our work demonstrates how AI can shift from a ``problem auditor'' to a ``solution co-creator'' by integrating multi-perspective dialogue with interactive repair, offering design implications for future creative tools.
\end{abstract}



\maketitle

\section{Introduction}
Contemporary design practice increasingly manifests as complex negotiations across disciplines. Designers are no longer mere ``visual craftsmen'' but have evolved into ``system coordinators'' who must balance user needs, business goals, and engineering constraints~\cite{huang2018understanding, ls2025challenges}. This elevation of their strategic role gives designers greater influence over product decisions, yet also introduces new tensions. For example, UX designers and product managers (PM) frequently disagree on product vision, user research priorities, and feature roadmaps~\cite{10.1145/3544548.3581273, bexiga2020closing}. As a result, designers must constantly switch between disciplinary mindsets—a process that imposes heavy cognitive load, weakens creative focus, and can even lead to professional burnout~\cite{kirschner2002cognitive, sweller1988cognitive}.

This growing complexity has spurred both academia and industry to explore automated tools aimed at alleviating designers’ burdens from repetitive evaluation tasks. Early tools primarily focused on single-dimension static checks, such as usability heuristics or WCAG accessibility compliance~\cite{duan2024generating, gao2024uianalyzer, lu2024ai, yang2021don}. However, by presenting feedback as mere ``problem lists,'' these tools disrupted designers’ creative flow~\cite{csikszentmihalyi1990flow} and disconnected problems from their potential solutions~\cite{duan2024uicrit, specai}. In recent years, some efforts have attempted to leverage large language models (LLMs) to improve critique processes—for instance, UICrit~\cite{duan2024uicrit} generates structured design issue reports. Yet overall, these systems remain confined to static, one-off checklist-style feedback, failing to deeply integrate with designers’ workflows or support the multidimensional trade-offs and ongoing negotiations required in authentic critiques.

This highlights a critical research gap: in practice, design critique is not only about identifying problems—it is a social process filled with negotiation, trade-offs, and co-creation~\cite{lee2018design}. Existing systems struggle to support this process for two main reasons: (1) they typically rely on single-perspective models, making it difficult to represent the conflicting stances and goals of internal stakeholders such as product managers, engineers, and designers; and (2) their feedback is largely static, lacking an interactive critique-to-action loop and failing to embed seamlessly into designers’ real workflows. Overcoming these limitations requires reframing AI’s role—from a static problem auditor to a dynamic solution co-creator~\cite{mark2008cost}—echoing broader calls within the HCI community to reimagine AI as a collaborative partner in creative work~\cite{chen2025genui, lee2025towards, liu2025human}.

To this end, we present \textit{CritiqueCrew}, a conversational critique tool integrated into Figma~\cite{figma2024}. CritiqueCrew connects problem discovery, explanation, and remediation into a seamless loop, enabling AI to participate throughout the process, from identifying issues to exploring solutions. It adopts two core design strategies: (1) \textbf{multi-perspective critique}, in which distinct expert modules approximate a cross-functional team to generate structured feedback; and (2) \textbf{empowering interactive remediation}, where in-context interactions such as canvas highlighting and real-time previews transform abstract feedback into concrete, actionable suggestions, while preserving designers’ full control over creative decisions.

We evaluated the effectiveness of CritiqueCrew through two controlled studies ($N = 48$). Study 1 benchmarked our system against a state-of-the-art tool, while Study 2 isolated the specific contribution of the multi-perspective architecture against a unified expert baseline. In this paper, we define \textit{design quality} along two outcome dimensions: (1) \textit{issue coverage} and (2) \textit{solution effectiveness} (expert ratings of visual aesthetics, usability \& compliance, and design integrity; see Section~\ref{sec:design-quality}). This definition is intentionally aligned with our multi-perspective critique framework: role-specialized perspectives are expected to broaden problem discovery (supporting higher \textit{issue coverage}), while cross-perspective synthesis helps form more coherent and actionable fixes (supporting higher \textit{solution effectiveness}). Results showed that CritiqueCrew significantly improved both issue coverage and solution effectiveness, while also enhancing designers' subjective usability ratings and trust in the tool.

The main contributions of this work are as follows:
\begin{itemize}
\item CritiqueCrew, a system that employs a novel approach of integrating multi-perspective dialogue with in-context interactive remediation. This approach transforms design critique into an actionable workflow that provides designers with diverse, cross-functional perspectives.

\item Results from two user studies ($N=48$) demonstrating the benefits of multi-perspective orchestration. Our findings show that CritiqueCrew significantly improves design quality and efficiency while reducing cognitive load, enhancing trust, and boosting perceived creativity support compared to unified expert models.

\item Key insights on shifting AI's role from a ``problem auditor'' to a ``solution co-creator,'' offering actionable design implications for the next generation of empowering creative support tools.
\end{itemize}

\section{Related Work}
\subsection{Human-AI Collaboration: From Tools to Partners}
Early HCI visions imagined computers not merely as tools, but as collaborators that augment human cognition \cite{licklider1960man}. Decades of research in interactive systems and Mixed-Initiative Interaction \cite{horvitz1999principles} further established the idea that humans and machines can flexibly share control, each contributing complementary strengths to a shared task. With recent advances in large language models (LLMs), this vision has become far more concrete: AI systems can now participate in ideation, generate alternative perspectives, and engage in back-and-forth reasoning, shifting their role from assistants to co-creative partners.

Co-creative systems have gained traction across writing, visual design, programming, and other creative domains, positioning AI as an active contributor rather than a passive generator \cite{moruzzi2024user, stray2025human, reza2025cowriting}. These systems help externalize reasoning, scaffold exploration, and broaden the user’s problem space. In design-oriented workflows, AI has also been used to bridge gaps between cross-functional roles—for instance, helping designers communicate intent to developers or stakeholders \cite{bexiga2020closing}. Across these lines of work, a common shift is emerging: AI is increasingly treated as a conversational partner capable of offering domain-relevant viewpoints.

Although prior systems have explored AI collaboration in creative tasks, far less attention has focused on how AI might participate in design critique—a process that extends beyond error detection to the shaping of interpretation and design direction. Rooted in art education \cite{feldman1994practical}, critique serves as a foundational practice that enables designers to clarify their creative intent and evaluate how their work is interpreted by others. As critique typically unfolds through conversational, small-group interactions \cite{luther2014crowdcrit}, it is inherently social and interpretive rather than mechanical. These properties make critique a compelling setting in which emerging AI systems, increasingly capable of dialogic and multi-perspective reasoning, may act not only as tools but as collaborative partners.

\subsection{Design Critique as a Social and Reflective Practice}
\label{sec:critique-social}
Design critique is widely recognized as a foundational mechanism for design iteration, distinct from error detection or performance optimization \cite{alabood2023systematic}. Schön characterizes design as a ``reflective conversation with the situation'' \cite{schon1986reflective}, wherein practitioners surface assumptions and reframe problems while interacting with evolving materials. Within this process, critique supports reflection-in-action by prompting designers to articulate the reasoning behind their choices and transform tacit judgments into explicit arguments \cite{blevis2007using, fischer1991role, stolterman2008nature}. Unlike validation-oriented procedures that seek binary compliance, critique helps designers navigate ambiguity and interpret conflicting constraints.

Critique is also fundamentally social and negotiative. Research in design cognition shows that evaluative discussions rely on the alignment of differing mental models and the exchange of rationale \cite{oh2013theoretical, shaffer2003portrait}. In professional settings, critique often involves cross-functional participants—such as designers, engineers, and product managers—who contribute distinct evaluative criteria. The interplay among these perspectives introduces productive tension that can reveal blind spots and stimulate strategic trade-offs \cite{bucciarelli1988ethnographic, saad2021ux, xu2015classroom}.

These theoretical perspectives suggest that computational support for critique must extend beyond static rule checking. Rather than serving as passive inspectors, effective systems should facilitate dialogic engagement and approximate the multi-perspective reasoning found in human teams. This distinction highlights the gap between conventional automated evaluation tools and the need for systems capable of supporting the interpretive, back-and-forth dynamics essential to high-quality design critique.

\subsection{Automated UI/UX Evaluation}
In contrast to the interpretive and multi-perspective nature of design critique described above, the domain of evaluation has historically focused on objective verification. Traditional methods, such as heuristic evaluation and usability testing \cite{jeffries1991user}, rely heavily on expert inspection to uncover flaws. While structured reviews can aid this process \cite{alabood2023systematic}, they remain resource-intensive and highly subjective, creating a demand for automated solutions that lower the barrier for novices \cite{lee2020guicomp}.

To scale evaluation efforts, researchers developed tools leveraging computer vision and machine learning to strictly validate designs against rules. Systems like UIS-Hunter \cite{yang2021don} and Nighthawk \cite{liu2023nighthawk} employ deep learning to automatically detect pixel-level ``design smells'' and GUI anomalies. Others, such as UIGuider \cite{yang2024uiguider} and UiAnalyzer \cite{gao2024uianalyzer}, assess the risk of non-conformance by constructing knowledge graphs. However, these approaches treat evaluation as a binary classification task, lacking the nuance to handle the ambiguous constraints typical of the reflection-in-action process.

More recently, the emergence of Vision-Language Models (VLMs) has forged new paths for automated evaluation by bridging visual perception with semantic reasoning. Research by Duan et al. \cite{duan2024generating} demonstrates that multimodal models can combine UI screenshots with design principles to generate detailed suggestions, while hybrid approaches integrate heuristic precision with VLM flexibility \cite{lu2024ai, yang2024uisgpt}. Beyond static analysis, researchers have also employed agents to simulate virtual user behavior for dynamic usability assessment \cite{xiang2024simuser}. Parallel efforts have focused on addressing the data scarcity problem; works like UICrit \cite{duan2024uicrit} establish large-scale datasets of professional feedback to enhance the critique capabilities of general-purpose VLMs. Despite these diverse advances, moving beyond surface-level visual analysis to a deeper, behavior-driven understanding of interface effectiveness remains a significant challenge \cite{jeon2025do}.

Ultimately, a fundamental gap persists: existing tools function as \textit{problem auditors} rather than \textit{critique partners}. Whether employing heuristics or VLMs, the output is overwhelmingly static—a list of issues or an annotated screenshot \cite{vera-amaro2025towards}. These systems lack the capacity for the \textbf{productive tension} and \textbf{cross-functional negotiation} defined in Section ~\ref{sec:critique-social}, forcing designers to manually bridge the divide between ``analysis'' and ``action.'' CritiqueCrew addresses this by situating evaluation within an interactive, multi-perspective dialogue that co-creates solutions.

\subsection{From Autonomous Simulation to Multi-Perspective Orchestration}

Recent advances in LLMs have renewed interest in agentic paradigms. A representative milestone is the Generative Agents framework \cite{park2023generative}, which integrates memory and planning to simulate social behavior. While effective for simulation, professional design workflows often require more structured coordination to ensure reliability.

Consequently, research has shifted toward functional role orchestration, assigning LLMs specialized personas to decompose tasks. In generative design, systems like PrototypeFlow \cite{yuan2024towards} and GUIDE \cite{kolthoff2025guide} use central controllers to coordinate modules for structured UI generation. Similarly, in evaluation, multi-perspective configurations simulate diverse stakeholders: PosterMate \cite{shin2025postermate} generates audience feedback, while PaperEval \cite{huang2025papereval} assigns verification roles for academic assessment. These works show that decomposing problems into functional roles improves outcomes.

However, most existing systems operate as one-way pipelines producing static outputs. They rarely support the negotiative dialogue and productive tension identified as essential for critique (Section ~\ref{sec:critique-social}). Professional critique demands more than simulating reactions; it requires the active orchestration of conflicting expert constraints into actionable decisions.

CritiqueCrew addresses this by orchestrating multi-perspective collaboration within a dialogic framework. Instead of relying on fully autonomous agents, we implement a structured Multi-Perspective Architecture where functional roles interact through mediated discourse. This design choice prioritizes interpretability over unconstrained autonomy, ensuring the system facilitates the rigorous reasoning of a cross-functional expert team.

\section{Formative Study}

To ensure that our system design is genuinely rooted in design practice workflows and core pain points, we conducted a formative study. Our goal was to deeply explore the difficulties faced by designers, product managers, and engineers during design critique and feedback within real product development processes, and, based on these insights, to investigate what role an AI design partner should play and what core capabilities it should possess.

\subsection{Methods and Participants}

We recruited and interviewed a total of 15 professionals working on the front line of product development. Following the method proposed by Francis et al. \cite{francis2010adequate}, we adopted an iterative data collection and analysis process with a stopping criterion based on theoretical saturation. We tracked the emergence of new codes after each interview. Detailed analysis of the first 12 participants established our core thematic framework, as no new first-level codes emerged regarding cross-functional conflicts after the twelfth interview. To ensure rigor, the subsequent three interviews served as confirmatory cases. Since these sessions validated existing themes without yielding additional conceptual insights, our findings report primarily focuses on the rich, distinct data provided by the initial 12 participants. The entire study protocol, including data collection from all 15 participants, was reviewed and approved by the University's Institutional Review Board (IRB). 

\textbf{Participants.} Among the 12 core participants, we included user interface (UI) designers (P1, P2, P3), interaction designers (P4, P5, P10), product managers (P6, P7, P11), and front-end engineers (P8, P9, P12). Their industry experience ranged from 1 to 10 years, and they served organizations of varying scales—from large technology companies to start-up teams. This diversity ensured that the collected data reflects common challenges across different organizational structures and product stages.

\textbf{Interview Procedure and Data Analysis.} We conducted semi-structured interviews lasting 30–45 minutes each. All interviews were held online and recorded/transcribed with participants’ consent. Interview discussions centered on two core topics:
\begin{itemize}
\item \textbf{Experiences}: Within an end-to-end design workflow, what authentic experiences and pain points did participants encounter during design critique?
\item \textbf{Needs}: After being introduced to the early concept of an ``AI Expert Review Panel,'' what functions and forms did they imagine and expect from an ideal AI collaborator?
\end{itemize}

We applied thematic analysis following the six-phase guide by Braun and Clarke \cite{braun2006using}. We adopted a hybrid inductive-deductive approach: while codes were initially generated inductively from the data, themes were refined deductively based on our research questions regarding ``Experiences'' and ``Needs.'' Two researchers independently conducted immersive readings of the core transcripts, generating initial codes. Through multiple rounds of discussion, these codes were iteratively refined aggregated into a stable codebook.
This codebook was then applied to the three confirmatory interviews (P13–P15) to verify thematic coverage. Regarding reliability, we prioritized negotiated consensus over statistical inter-rater reliability metrics (e.g., Cohen’s Kappa), aligning with qualitative research standards in HCI \cite{mcdonald2019reliability}. Any discrepancies were resolved through discussion, ultimately yielding three core themes spanning the dataset. The complete codebook, including definitions and representative quotes for all themes, is detailed in Appendix~\ref{sec:appendix_codebook}.

\subsection{Findings}
\subsubsection{\textit{Beyond UI Boundaries—Designers Seek Proactive, Multi-Faceted Expert Feedback to Navigate Cross-Functional Conflicts.}}  
Our interviews consistently revealed that design critique extends far beyond aesthetic or usability evaluations. Instead, it functions as a primary venue for negotiating and reconciling competing demands from cross-functional stakeholders, particularly from product management and engineering. While this negotiation is crucial, it is also substantially time-consuming. As an interaction designer (P5) described, a single comprehensive critique cycle—spanning from internal reviews to achieving cross-team alignment—\textit{``could take about five hours.''}  This significant time investment reflects the labor required to bridge the inherent tensions among different disciplinary perspectives. Situated at the nexus of these competing constraints, designers are tasked with finding optimal solutions. This position underscores a critical need for proactive and integrated expert feedback early in the design process to anticipate and mitigate such conflicts. This need was particularly salient in the following two scenarios of tension:

\begin{itemize}
\item \textbf{Balancing engineering feasibility and design innovation}: A prominent tension emerged between engineering pragmatism and design innovation. A front-end engineer (P8) recounted an instance where a designer proposed using a non-standard font to achieve a distinctive visual treatment. From the designer's standpoint, this was a pursuit of visual innovation aimed at enhancing the user experience. However, the engineer countered that while technically implementable, this deviation would violate established coding conventions and introduce significant technical debt, complicating future maintenance. This case highlights how designers' responsibilities extend beyond ideation to encompass navigating the complex trade-offs between novel design concepts and the long-term scalability and stability of the engineering architecture.
\item \textbf{Negotiating between business demands and design system guidelines}: Another recurring conflict involved mediating between immediate business objectives and long-term design consistency. A UI designer (P2) described a negotiation with a product manager who insisted on placing a critical business feature in the \textit{``most prominent position at the top of the interface.'' } For the designer, this directive not only contravened the established design system guidelines but also necessitated the creation of a non-standard component, thereby compromising product coherence. Designers found themselves balancing short-term business goals against the long-term integrity of the user experience. This conflict, as P2 noted, forced the team to \textit{``spend a lot of time in multiple meetings'' } to forge a compromise.
\end{itemize}

Taken together, these recurring tensions demonstrate that designers act as key conflict mediators in collaborative projects. They need more than isolated, unidirectional feedback—they need a virtual expert advisory panel that can simulate realistic decision-making scenarios and integrate perspectives from engineering, product vision, and user experience. Designers are eager to anticipate, surface, and reconcile potential conflicts early, reallocating team resources from costly late-stage rework to high-value creative innovation.

\subsubsection{\textit{Beyond Problem Lists—Designers Need Contextualized Solutions to Drive Iteration}}  

Our findings indicate that feedback which only identifies problems is perceived as insufficient. As multiple designers emphasized, the most valuable critique not only diagnoses issues but also provides actionable pathways toward solutions. This was clearly articulated by a senior designer (P3), who defined ideal feedback as \textit{``Tell me where the problem is, and give me a few feasible options to fix it.''}  She recounted an instance where a vague critique of her design being \textit{``too messy''} was made concrete when her lead identified the root cause (excessive icon colors) and offered specific remedies. This exemplifies an effective feedback structure that we characterize as a three-part process: diagnosis (the \textit{``what''}), localization (the \textit{``where''}), and recommendation (the \textit{``why''}).

This desire for a complete feedback loop extends to designers' expectations for AI tools. A junior designer (P10) envisioned an ideal tool as one that goes beyond flagging contrast errors to proactively \textit{``recommend several WCAG-compliant safe color values.''} The underlying goal is to offload the tedious work of trial-and-error and accelerate creative problem-solving. Thus, for an AI partner to be truly effective, it cannot be a passive critic. It must act as a collaborative problem-solver, providing a scaffolded pathway from issue detection to creative resolution by delivering not just data, but decisive, implementable suggestions.

\subsubsection{\textit{Beyond Auditing Tools—Designers Seek Empowering and Trustworthy Partnerships to Support Creative Autonomy}}  

Discussions of human-AI collaboration were dominated by the intertwined themes of trust and control. Participants emphasized that their adoption of an AI tool hinges on its ability to foster a partnership that augments, rather than dictates, their creative process.

Building this trust rests on two pillars. The first is \textit{transparency and explainability}~\cite{afroogh2024trust}. As interaction designer P5 articulated, trust is not instantaneous but progressively established. He emphasized that trust is cultivated when an AI can justify its recommendations, stating, \textit{``If the AI can justify its recommendations, that’s what builds my trust.''} This underscores the critical need for designers to understand the underlying rationale behind AI suggestions. Consequently, a \textit{``black-box''} system with opaque reasoning is unlikely to gain acceptance. The second pillar is \textit{reliability}~\cite{afroogh2024trust}. P3 noted that when an AI misinterprets prompts or generates irrelevant outputs, often described as \textit{AI hallucinations}~\cite{ji2023hallucination}, its credibility is immediately compromised. Without consistent and reliable performance, trust cannot be sustained.

Beyond trust, designers were emphasized about retaining control and professional autonomy. This principle manifested as a demand for empowerment, not replacement, with designers insisting on final decision-making authority. As P3 aptly put it, the ideal AI is \textit{``An advisor who doesn’t get in the way of my decisions.''}  Operationally, this translates to an AI that furnishes diverse options, clarifies trade-offs, and enables exploration, while always reserving the final judgment for the human designer. In essence, designers seek a reliable and transparent collaborator that respects their autonomy and enhances their capabilities, rather than an automated auditor that diminishes their authority.

\section{System Design}

Our formative study revealed designers’ deep desire for multi-dimensional insights, contextualized remediation, and empowering partnerships when collaborating with automated tools. To systematically respond to these expectations, we developed CritiqueCrew, a Figma-integrated tool designed to shift AI from a mere ``problem auditor'' to a ``solution co-creator.'' This section first introduces the system’s design goals and overall architecture, then elaborates on its core multi-perspective critique model, and finally illustrates how users collaborate with the system through a complete interaction workflow.

\subsection{Design Goals}

Building on the three core findings from our formative study, we established three design goals for CritiqueCrew:

\textbf{DG1: Incorporating multi-perspective, expert-level critique.} Instead of relying on a single viewpoint, CritiqueCrew orchestrates distinct functional roles to generate feedback that integrates constraints from user experience, product vision, and engineering feasibility.

\textbf{DG2: Seamlessly translating critique into actionable solutions.} Rather than delivering abstract suggestions, CritiqueCrew provides interactive remediation tools that allow designers to preview and apply modifications directly within their design context.

\textbf{DG3: Fostering an empowering and trusting partnership.} To build trust, interactions are designed to be conversational and explainable. The system explicitly articulates trade-offs and rationale while ensuring designers remain the final decision-makers.

\subsection{System Architecture}
\begin{figure*}[t]
  \centering
  \includegraphics[width=\textwidth]{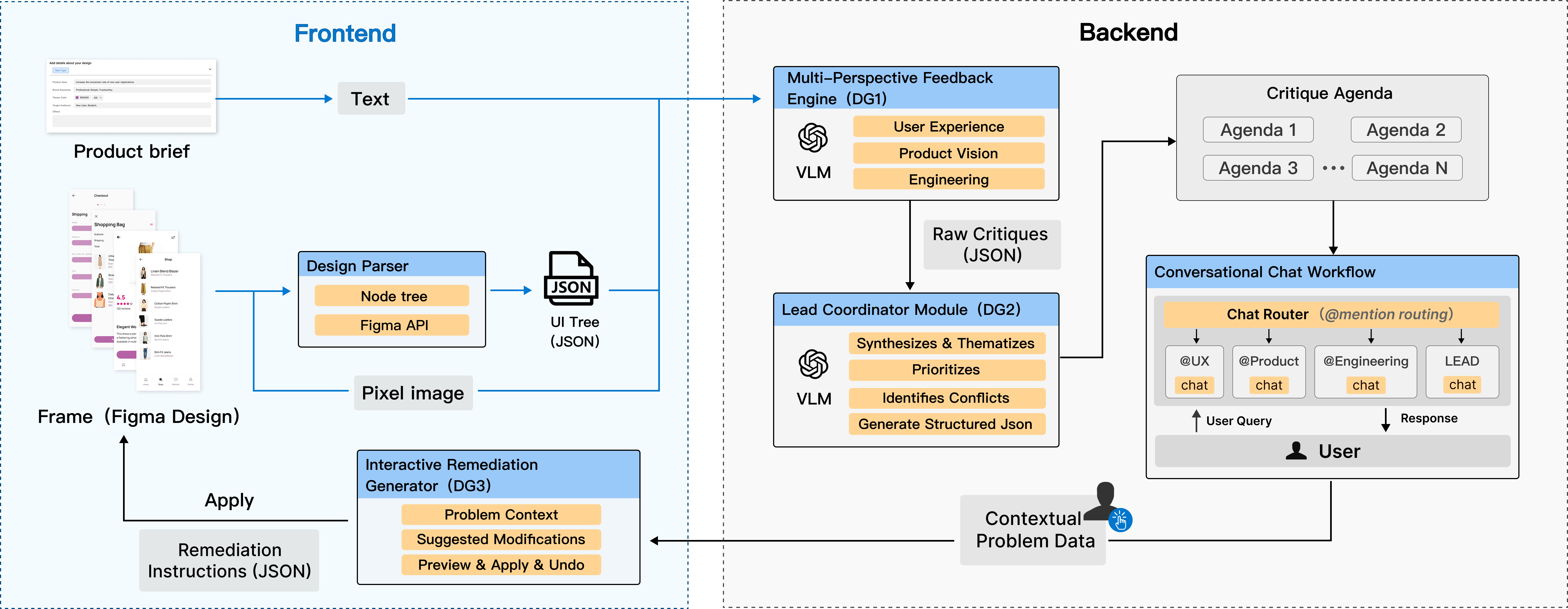}
  \caption{System architecture of CritiqueCrew}
  \Description{The figure shows the overall system architecture of CritiqueCrew. The workflow proceeds from left to right. (Left) Front-end Inputs: User inputs include a product brief (contextual goals) and a Figma frame (visual design). (Middle) Backend Processing: The Design Parser converts the Figma node tree and pixel image into structured JSON. This data feeds into the Multi-Perspective Feedback Engine, where parallel Expert Modules evaluate the design through distinct lenses (UX, PM, Engineer). The Lead Coordinator Module then synthesizes these raw insights into a structured Critique Agenda, prioritizing issues and identifying conflicts. It also manages a Chat Router that directs context-specific queries (e.g., @UX) to the corresponding module during dialogue. (Right) Interactive Remediation: The Interactive Remediation Generator translates feedback into actionable JSON instructions. Finally, these critiques and remediation options are returned to the front-end plugin, enabling users to preview and apply changes directly on the Figma canvas.}
  \label{fig:architecture}
\end{figure*}

CritiqueCrew follows a client–server architecture designed to tightly integrate advanced analysis capabilities with a seamless front-end experience. The overall system architecture is shown in Figure~\ref{fig:architecture}.

The frontend is a lightweight Figma plugin built with React~\cite{react_github}. It serves as the main interface for user interaction, responsible for rendering the critique agenda and a set of interactive remediation tools. Through the Figma Plugin API, the frontend can read design layer data from the canvas in real time and write user actions (such as highlighting elements or updating properties) back onto the canvas, ensuring a fluid workflow.

The backend adopts a modular design, consisting of four core components that orchestrate the critique process:

\begin{itemize}
\item \textbf{Design Parser}: Converts Figma’s complex native node tree into a structured JSON format suitable for LLM processing. This JSON not only preserves the hierarchy of the interface but also integrates semantic and visual attributes (such as text labels, positions, sizes, and colors), making raw design data interpretable for downstream analysis. Figure~\ref{fig:json} illustrates a simple frame and its JSON representation.

\begin{figure}[t]
  \centering
  \includegraphics[width=0.65\linewidth]{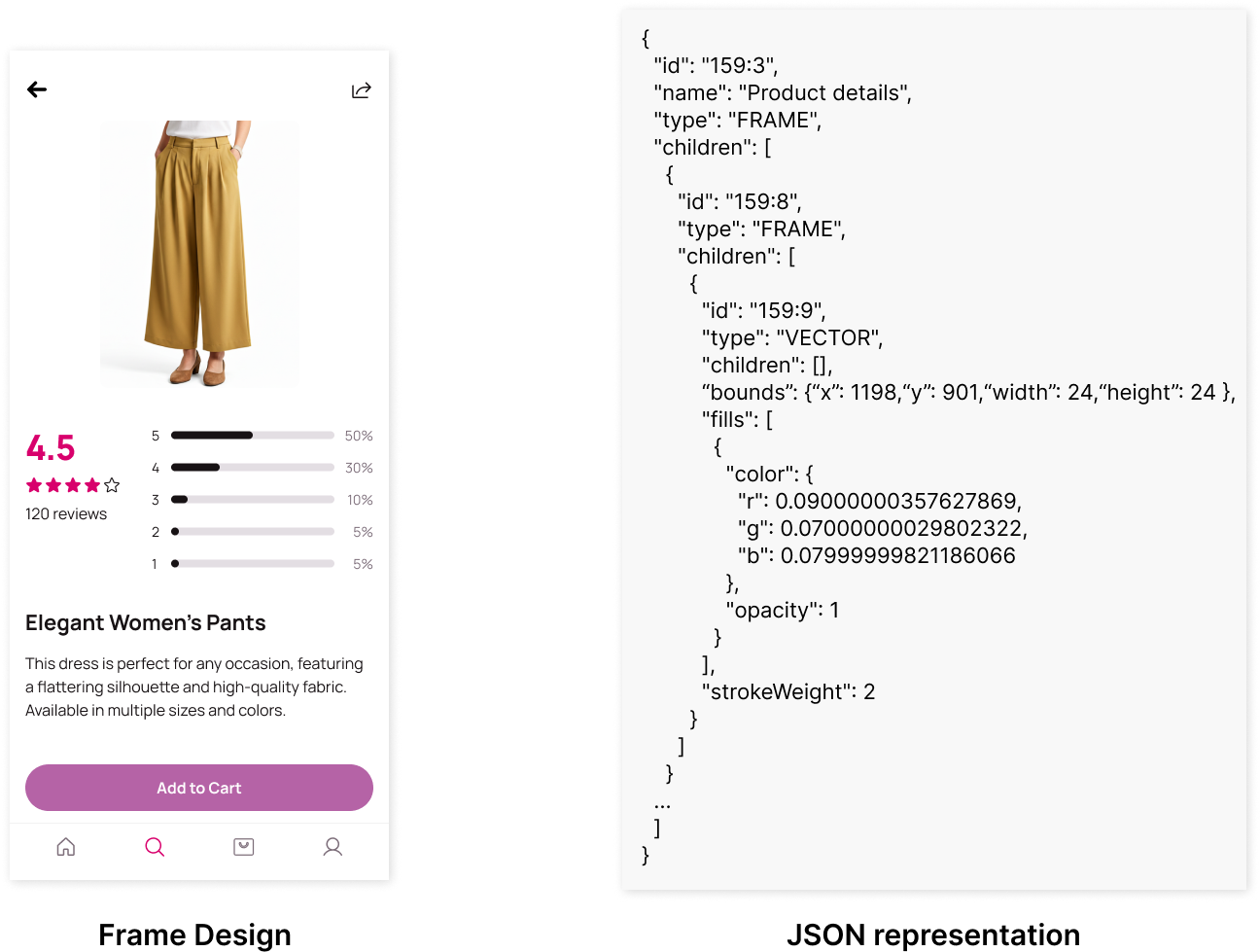}
  \caption{Frame Design and JSON Representation}
  \Description{The figure shows a mobile product detail page on the left and its parsed JSON representation on the right. The left side displays a Figma frame containing an image, ratings, text descriptions, and an action button, representing a typical UI layout. The right side shows the corresponding JSON structure produced by the Design Parser, with hierarchical FRAME and VECTOR nodes, each containing properties such as bounds, fills, and stroke information. The figure illustrates how the visual design is translated into a structured machine-readable format.}
  \label{fig:json}
\end{figure}

\item \textbf{Multi-Perspective Feedback Engine}: The core of multi-dimensional critique (DG1). Implemented using the LangGraph~\cite{langgraph} framework to ensure deterministic execution flow, this engine instantiates parallel Expert Modules powered by OpenAI’s GPT-4o~\cite{openai2024gpt4o}. Each module evaluates the parsed design data through a distinct functional lens (e.g., user experience, product vision, engineering feasibility), generating specialized critiques that are later aggregated into a comprehensive review.

\item \textbf{Lead Coordinator Module}: Acts as the ``orchestration hub'' of the system, designed to empower designers with coherent guidance (DG3). It synthesizes raw feedback from different expert modules, organizes them into themes, prioritizes issues, and explicitly highlights conflicts or trade-offs. This process transforms scattered insights into structured, traceable decision support. Additionally, this module manages a lightweight routing mechanism that directs context-specific queries (e.g., @UX, @PM, @Engineer) to the corresponding expert module during interactive dialogue, while serving as the default responder for general inquiries.

\item \textbf{Interactive Remediation Generator}: The key bridge between critique and action (DG2). Based on the context of each identified issue, this component dynamically generates remediation suggestions and produces the UI code required by the frontend. Designers can then preview, apply, or undo modifications directly on the canvas, turning abstract feedback into immediate solutions without disrupting their workflow.
\end{itemize}

\subsection{Multi-Perspective Critique Model}

To achieve our design goals, we developed a \textbf{Multi-Perspective Critique Model} composed of specialized expert modules that approximate the reasoning of a cross-functional review team, alongside a Lead Coordinator responsible for synthesis and empowerment. To ensure each module adheres to its specific functional lens, we designed rigorous system prompts (detailed in Appendix \ref{sec:prompts}).

\subsubsection{Expert Modules: Approximating Cross-Functional Perspectives}

To realize DG1, our model orchestrates a ``virtual review panel'' consisting of three core Expert Modules. These modules are designed to emulate the diverse evaluative criteria found in real-world product development, providing feedback that balances depth and breadth:

\begin{itemize}
\item \textbf{User Experience Module}: \textit{The guardian of the user}. This module synthesizes the roles of user researcher, UI designer, and accessibility expert.  It evaluates usability, information architecture, consistency, and accessibility against industry benchmarks such as Nielsen’s usability heuristics and WCAG standards. Its guiding question is: ``Is this design clear, efficient, and accessible to all users?''

\item \textbf{Product Vision Module}: \textit{The navigator of goals}. This module represents the strategic perspective of product managers and business stakeholders. Designers can provide product goals (e.g., increase new user registration conversion) or brand keywords (e.g., professional, minimal). Based on this context, it applies zero-shot reasoning to evaluate whether the design aligns with business objectives, product strategy, and brand identity. Its guiding question is: ``Does this design effectively drive business goals and faithfully communicate our brand values?''

\item \textbf{Engineering Module}: \textit{The pragmatic engineer}. This module embodies the constraints of a front-end developer, acting as a technical gatekeeper. It focuses on assessing engineering feasibility, estimating implementation cost, and identifying potential performance bottlenecks. Its guiding question is:  ``Can this design be implemented efficiently with the current tech stack, and what are the cost and performance implications?''
\end{itemize}

To ensure consistency and interoperability, the output of each module follows a unified structured format: \texttt{issue\_id}, \texttt{source\_role}, \texttt{component\_target} (linking to the Figma layer ID), \texttt{severity}, \allowbreak\texttt{critique\_text}, \texttt{rationale}, \texttt{suggestion}. This structured representation makes AI-generated insights computable, sortable, and directly actionable.

\subsubsection{Lead Coordinator: From Auditor to Enabler}

To achieve DG3, we introduce a Lead Coordinator Module to orchestrate the entire critique process. Its primary mission is not to make final judgments, but to empower designers as the ultimate decision-makers through three key steps:
\begin{itemize}
\item \textbf{Synthesize \& Thematize}: Aggregates raw feedback from different expert modules according to related UI components. This process transforms what might otherwise be a scattered list of critiques into a structured, navigable agenda, effectively reducing designers’ cognitive load.

\item \textbf{Prioritize}: Orders the agenda based on issue severity. For example, a critical accessibility barrier (such as a WCAG AA violation) is prioritized over a minor aesthetic suggestion, guiding designers to address the most impactful issues first.

\item \textbf{Surface Conflicts \& Trade-offs}: When recommendations diverge—for instance, the Product Module suggesting a brand color for conversion, while the UX Module flags its insufficient contrast—the Lead Coordinator does not impose a correct solution. Instead, it explicitly presents the trade-off (e.g., Visual Prominence vs. Accessibility) and frames the decision for the designer, ensuring the tension is productive rather than paralyzing.
\end{itemize}

\subsection{Interaction Workflow}

\begin{figure*}[t]
  \centering
  \includegraphics[width=\textwidth]{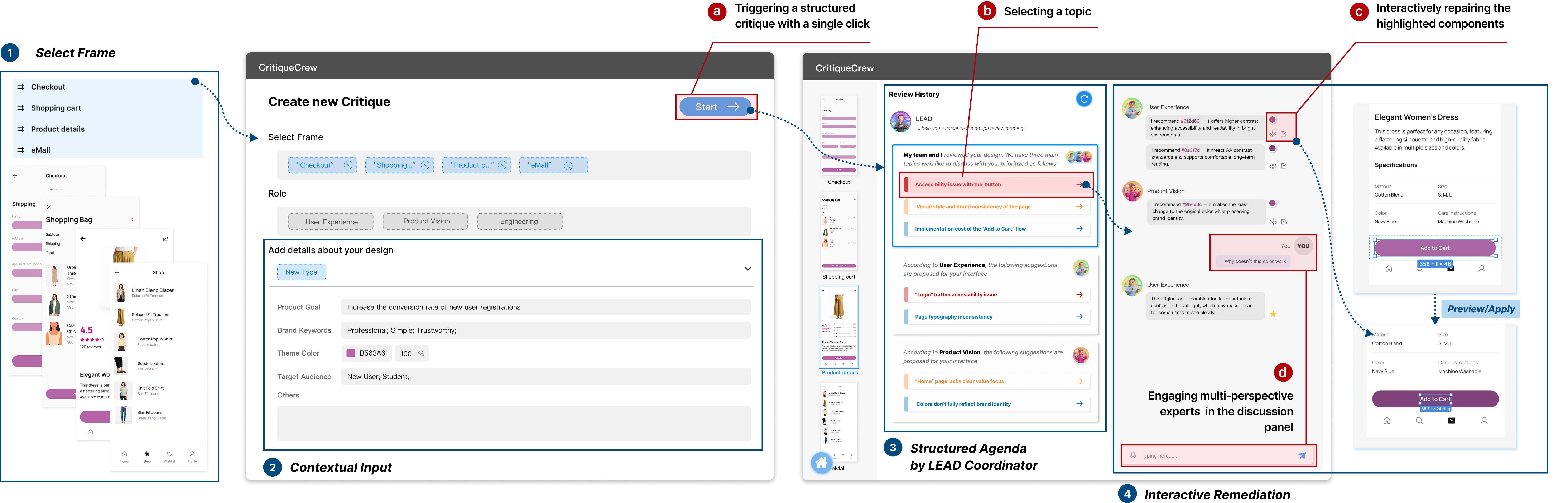}
  \caption{Interaction workflow}
  \Description{The figure displays the CritiqueCrew interface divided into three main interaction stages. First, on the left, a "Create new Critique" panel allows users to select Figma frames (e.g., Checkout, Shopping Bag) and input design context, such as Product Goal, Brand Keywords, and Theme Color. A blue "Start" button initiates the process. Second, in the center, the main "Review History" panel shows a critique agenda generated by the Lead Coordinator. The agenda lists prioritized issues, such as "Accessibility issue with the button," followed by specific comments from Expert Roles (labeled as User Experience, Product Vision). Below the agenda, a chat interface displays a conversation where the user asks, "Why doesn't this color work?" and the User Experience role responds with a rationale about contrast standards. Third, on the right, a callout shows the "Interactive Remediation" feature. When a user selects a specific color suggestion (e.g., changing a button to purple), the corresponding button on the Figma canvas is visually highlighted. A small popup allows the user to preview the change and apply it directly with a single click. Overall, the interface demonstrates how users configure critique parameters, review synthesized multi-perspective feedback, discuss conflicts via chat, and apply fixes directly to the design.}
  \label{fig:interaction}
\end{figure*}

The core interaction workflow of CritiqueCrew is designed as a seamless loop of 
``invoke--analyze--interact--resolve,'' transforming design critique from a one-off event into an ongoing dialogue embedded in the creative process. The interaction workflow is illustrated in Figure~\ref{fig:interaction}.

\begin{enumerate}
  \item \textbf{Invoking Critique and Reviewing the Agenda}: 
  When designers complete a draft or a milestone in Figma, they can invoke CritiqueCrew with a single click to start a critique session (Figure~\ref{fig:interaction} \raisebox{-0.28ex}{\includegraphics[height=1.2em]{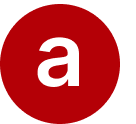}}). The system immediately collects the selected frame’s layer data and sends it to the backend expert modules for parallel analysis. Within seconds, the Lead Coordinator synthesizes the results and presents a structured 
  critique agenda in the sidebar. This agenda not only provides a clear roadmap for subsequent improvements but also establishes a multi-dimensional, structured dialogue from the outset (DG1), laying the foundation for building trust (DG3).

  \item \textbf{Contextualizing Feedback with Interactive Remediation}: 
  Designers can then select any issue from the agenda to begin working on it (Figure~\ref{fig:interaction} \raisebox{-0.3ex}{\includegraphics[height=1.2em]{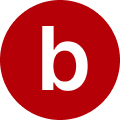}}). Once selected, the system creates a tight link between the canvas and the feedback: the Figma canvas automatically zooms in and highlights the corresponding UI component, while the sidebar switches to display detailed feedback, theoretical rationale, and context-sensitive interactive remediation 
  tools (Figure~\ref{fig:interaction} \raisebox{-0.3ex}{\includegraphics[height=1.2em]{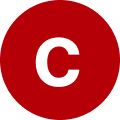}}). For instance, if a color contrast issue is detected, the system provides a palette that complies with WCAG standards. Designers can preview the changes instantly and apply them with a single click, allowing problem localization, solution exploration, and design 
  execution to be seamlessly coupled without disrupting their creative flow (DG2).

  \item \textbf{Resolving Conflicts through Multi-Perspective Dialogue}: 
  For complex issues that involve trade-offs, designers can open a dedicated dialogue panel for the topic (Figure~\ref{fig:interaction}
  \raisebox{-0.28ex}{\includegraphics[height=1.2em]{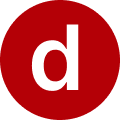}}). Here, the system not only presents the final recommendations from 
  different functional roles but also summarizes their divergent perspectives, making designers feel as though they are part of a transparent review meeting where diverse perspectives are surfaced. Moreover, designers 
  can use the \texttt{@} feature to directly query a specific role, such as 
  ``\texttt{@Engineer why is this animation costly to implement?}'' This mechanism transforms 
  AI’s traditional ``black-box'' reasoning into a transparent, participatory deliberation, revealing the underlying tensions in design decisions and empowering designers to make the final judgment while strengthening a partnership of trust (DG3).
\end{enumerate}

\section{Evaluation}

We adopted a sequential evaluation strategy. Study 1 assesses the holistic effectiveness of CritiqueCrew against state-of-the-practice industry tools. Motivated by the significant performance gains observed, we then designed and conducted Study 2—with an independent group of participants and new design tasks—to deconstruct the source of these improvements, specifically isolating the contribution of our multi-perspective architecture from the interaction modalities. This two-stage approach addresses the following research questions:

\begin{enumerate}
  \item \textbf{RQ1: Design Quality} \\
  How does the full CritiqueCrew system compare to state-of-the-art tools, and does the \textbf{orchestration of specialized expert roles} specifically yield better outcomes than a single generalist model?

  \item \textbf{RQ2: User Experience} \\
  How does CritiqueCrew’s conversational workflow influence designers’ subjective experience, including usability, cognitive load, trust and perceived creativity support?

  \item \textbf{RQ3: Interaction Patterns} \\
  Compared to static analysis tools, how do designers perceive and make use of critiques from different functional roles within CritiqueCrew, especially when their perspectives conflict?
\end{enumerate}

\subsection{Study 1: CritiqueCrew vs. Baseline (SpecAI)}

Study 1 benchmarks CritiqueCrew against SpecAI, a representative static evaluation plugin. This comparison assesses the practical value of our full system and its core conversational paradigm.

\subsubsection{Study Design}
We adopted a within-subjects design, where each participant completed a design optimization task under two different conditions. This design effectively controls for individual differences (e.g., design experience).

\textbf{Condition 1: Baseline (SpecAI).}  
To compare CritiqueCrew with current state-of-the-art automated evaluation tools, we selected a real Figma community plugin called \textit{SpecAI}~\cite{specai} as the baseline. This plugin uses AI to improve design quality, accessibility, and copywriting, and presents its feedback in the sidebar as a static list of issues. Such an interaction pattern faithfully represents the ``auditor-style'' tools we aim to improve upon. At the time of our study, SpecAI had about 2,400 users, making it representative of advanced static analysis tools currently in practice.

\textbf{Condition 2: CritiqueCrew.}  
In this condition, participants used the full CritiqueCrew system. They engaged in dialogue with multiple expert modules and explored modifications through interactions such as canvas highlighting and real-time preview. Both tools were configured to detect the same categories of issues to ensure a fair comparison.

\subsubsection{Participants}
We conducted an a priori power analysis (G*Power) assuming a large effect size ($d_z = 0.8$, $\alpha = .05$, $1 - \beta = .80$), consistent with previous evaluations of AI-assisted design tools. This indicated a minimum sample of 15. To ensure stable estimates, we aimed for at least 24 participants per study.

We successfully recruited 24 professional UI designers (15 female, 9 male) from online design communities for Study 1. Participants had an average age of 26.8 years ($SD = 1.77$) and an average of 3.5 years ($SD = 0.9$) of professional design experience. Recruitment was conducted via professional networks, and no participants overlapped between the two studies. Each participant received a gift card as compensation. The study protocol was reviewed and approved by the university’s Institutional Review Board (IRB).

\subsubsection{Task \& Materials}
We prepared two multi-frame Figma design mockups for the experiment: a course detail page for an online education platform and a checkout flow for an e-commerce application. To represent realistic design artifacts, each mockup was composed of a sequence of 4-5 frames and was assigned to one of the two experimental conditions. Both mockups were seeded with a similar number of pre-defined issues (approximately 16). These issues were intentionally designed to cover the core focus areas of our system’s three \textbf{perspectives} (user experience, product vision, and engineering feasibility) and included several potential conflicts to evaluate how designers make decisions when faced with trade-offs.

To ensure task equivalence, we conducted a pilot study with two external designers, who confirmed that the cognitive effort and time required to complete both tasks were comparable. During the study, participants were allotted 20 minutes per condition to identify and resolve as many issues as possible within the assigned mockup using the provided tool. The order of the tasks and conditions was counterbalanced using a Latin Square design to mitigate order effects.

\subsubsection{Procedure}
The entire study lasted approximately 80 minutes and was conducted online via video conferencing. The procedure consisted of four parts: an introduction and consent process (5 minutes), tutorials and formal tasks for both tools (30 minutes each), and a final semi-structured interview (15 minutes). During the tasks, participants were encouraged to think aloud. The interviews focused on their comparative experiences with the two tools and their decision-making processes.

\subsection{Study 2: Role Specialization Analysis}

While Study 1 demonstrated the superiority of CritiqueCrew over static analysis tools, this comparison bundled multiple variables (backend architecture + conversational interface). To rigorously identify the active ingredient of our success, Study 2 isolates the role orchestration mechanism by controlling for all interaction affordances.

\subsubsection{Study Design}

Study 2 utilized a strictly independent sample of participants (none participated in Study 1). To ensure the findings were not artifacts of specific design files while maintaining comparability in domain complexity, we developed two \textbf{new} multi-frame Figma mockups. These mockups retained the same industry themes as Study 1 (Online Education and E-commerce) but featured completely distinct content and layout structures. Each mockup was seeded with approximately 16 issues matched in difficulty. The study employed a within-subjects design where each participant completed both tasks, one with the Unified Expert and one with the Multi-Perspective system, with the order counterbalanced.

\textbf{Condition 1: CritiqueCrew (Unified Expert).}
In this condition, the backend was configured to run a single GPT-4o instance prompted as a ``comprehensive design expert.'' The model was provided with a merged guideline that combined criteria for user experience, product vision, and engineering feasibility into a single system prompt. The front-end interaction remained identical across conditions. Unlike the multi-perspective setup, critique generation was not role-separated, and aggregation and prioritization were implicitly handled within the unified expert’s single response rather than through a separate, explicit coordination layer.

\textbf{Condition 2: CritiqueCrew (Multi-Perspective).}
In this condition, participants used the full system with explicit UX Designer, Product Manager, and Engineer roles orchestrated by the Lead Coordinator. Critiques were generated by distinct expert modules, enabling explicit conflict identification and trade-off reasoning. The Lead Coordinator makes synthesis and prioritization explicit by consolidating cross-role feedback and surfacing conflicts.

\subsubsection{Participants}
Study 2 involved a new group of 24 professional UI designers (11 female, 13 male). As noted, this sample was completely independent from Study 1. Their demographic profile was comparable: average age 27.5 years ($SD = 2.1$) and 3.2 years ($SD = 1.1$) of professional experience. All participants received standard compensation, and the study protocol was reviewed and approved by the university’s ethics board.

\subsection{Data Collection and Analysis}
\label{sec:design-quality}
We employed a mixed-methods approach, collecting both quantitative and qualitative data.

\textbf{Quantitative Data.} 
We operationalized design quality using a multi-dimensional assessment framework:

\begin{itemize}

  \item \textbf{Issue Coverage}: The number of predefined issues successfully resolved by participants. Given the fixed task duration, this measure also served as a proxy for efficiency, reflecting participants’ output within the allotted time.

  \item \textbf{Solution Effectiveness}: Three senior design experts, blind to the experimental conditions, independently evaluated all anonymized final designs. Guided by a structured quality model informed by recent multi-dimensional UI evaluation metrics \cite{duan2024uicrit,zen2023quality,xie2025datawink}, experts assessed three dimensions: \textbf{Visual Aesthetics}, \textbf{Usability \& Compliance}, and \textbf{Design Integrity}. They assigned a single 7-point rating (1 = very poor, 7 = excellent) to the effectiveness of the solution.
  
  \item \textbf{Subjective Questionnaires}: We collected scores from the System Usability Scale (SUS)(Study 1 only), the NASA-TLX cognitive workload scale, and the Trust in AI scale (TAI)~\cite{hoffman2023measures}.For Study 2, we additionally included the Creativity Support Index (CSI)\cite{cherry2014quantifying} to support creative exploration.
\end{itemize}

For statistical analysis, we first conducted Shapiro–Wilk tests to assess the normality of paired difference scores. When normality assumptions were met, we used paired-sample t-tests; otherwise, we used Wilcoxon signed-rank tests. We report test statistics, p-values, and corresponding effect sizes (Cohen’s $d$ for t-tests; rank-biserial $r$ for Wilcoxon tests).

\textbf{Qualitative Data.}
Qualitative data included transcripts of think-aloud sessions and interviews. Two researchers coded the transcripts using thematic analysis. The focus was on understanding how designers interpreted and balanced the feedback from different functional roles, and how the structured dialogue influenced their decision-making.

\section{Results}

In this section, we present the findings from our user study to answer our three core research questions. We first report the quantitative comparison results regarding design quality (RQ1) and subjective user experience (RQ2), followed by a deep dive into the qualitative insights on designers’ interaction patterns with CritiqueCrew (RQ3).

\subsection{RQ1: CritiqueCrew Improves Design Quality and Efficiency}

\begin{figure}[t]
\centering
\includegraphics[width=\columnwidth]{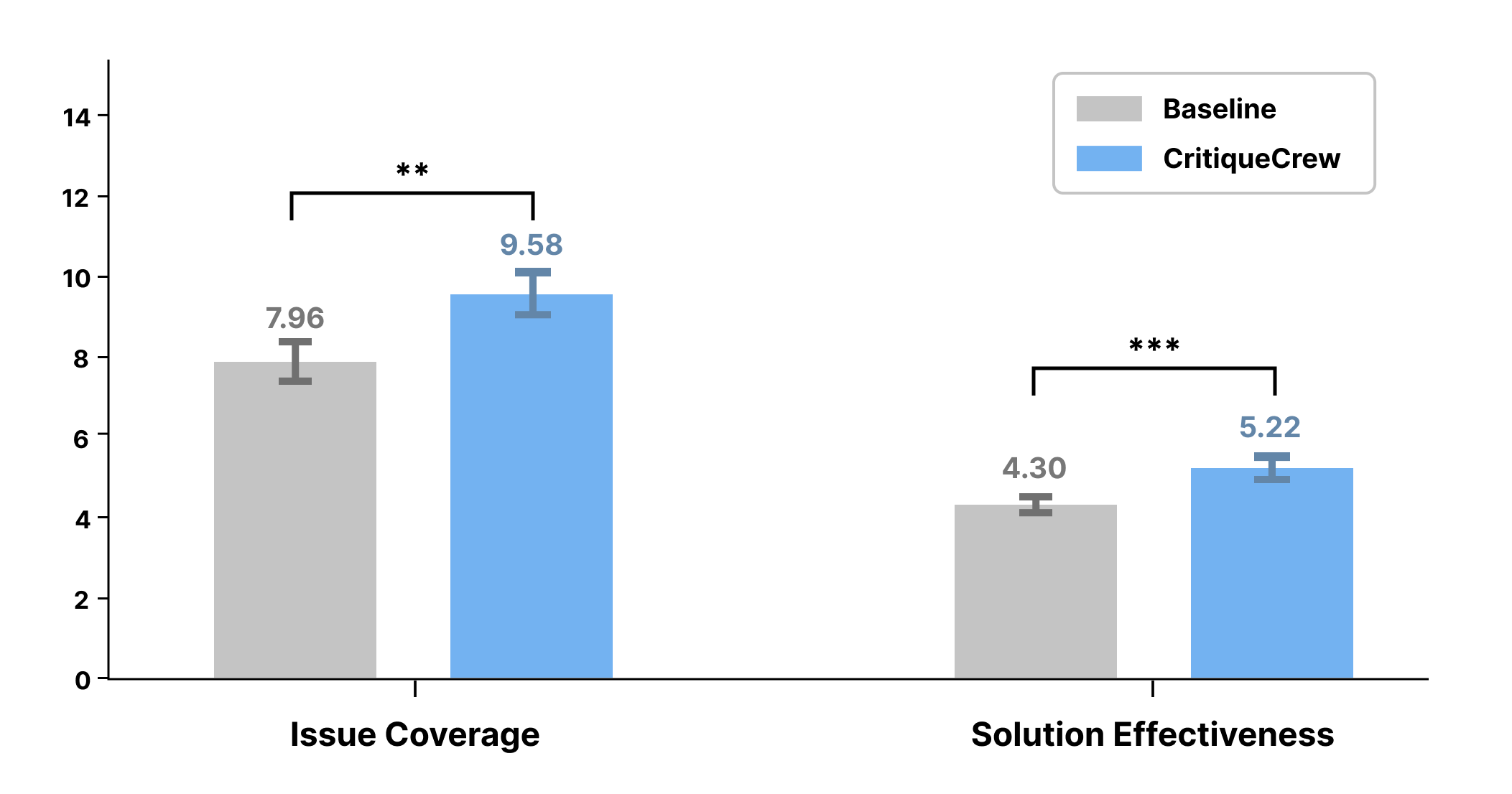} 
\caption{Comparison of design outcome quality between CritiqueCrew and the baseline tool in the realistic design task. CritiqueCrew led to broader issue coverage (left) and significantly higher solution effectiveness (right) than the baseline tool. Error bars represent standard errors. **$p < .01$, ***$p < .001$.}
\Description{Two bar charts comparing Baseline and CritiqueCrew with exact data labels. The left chart shows 'Issue Coverage', where CritiqueCrew (9.58) scores significantly higher than Baseline (7.96). The right chart shows 'Solution Effectiveness', where CritiqueCrew (5.22) is rated significantly higher than Baseline (4.30). Significance markers indicate p < .01 for coverage and p < .001 for effectiveness.}
\label{fig:quality1}
\end{figure}

\begin{figure}[t]
\centering
\includegraphics[width=\columnwidth]{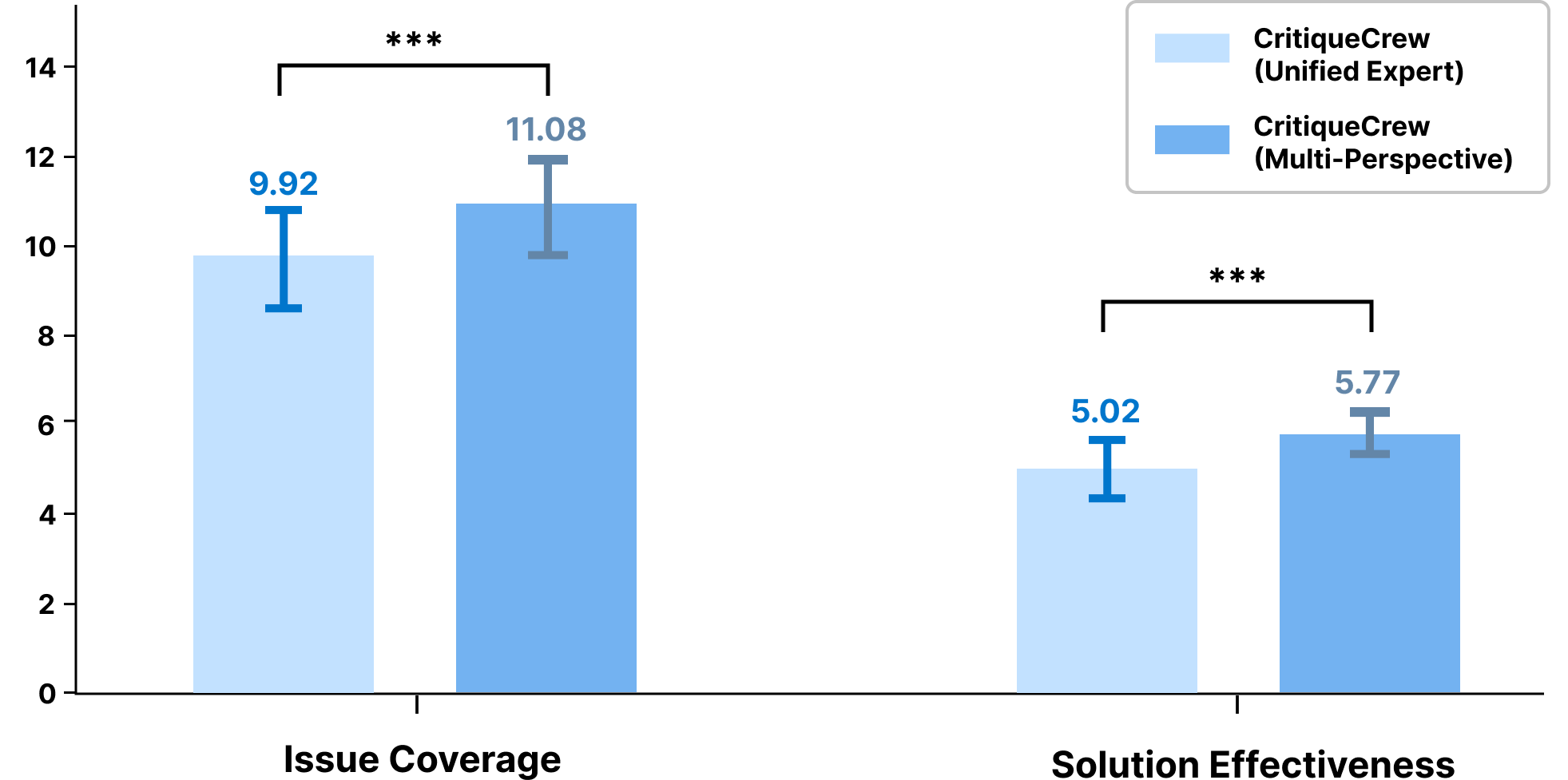} 
\caption{Comparison of design outcome quality between the two CritiqueCrew modes: Unified Expert vs. Multi-Perspective. The Multi-Perspective mode yielded significantly higher issue coverage (left) and solution effectiveness (right), confirming the benefits of role-based critique decomposition. ***$p < .001$.}
\Description{Two bar charts comparing Unified Expert and Multi-Perspective modes with exact data labels. The left chart displays 'Issue Coverage', showing that the Multi-Perspective mode (11.08) identifies more issues than the Unified Expert mode (9.92). The right chart displays 'Solution Effectiveness', showing that designs from the Multi-Perspective mode (5.77) are rated higher than those from the Unified Expert mode (5.02). Both differences are statistically significant with p < .001.}
\label{fig:quality2}
\end{figure}

Our first research question (RQ1) investigates whether CritiqueCrew leads to better design outcomes. We report results across two key metrics: \textbf{Issue Coverage} (efficiency) and \textbf{Solution Effectiveness} (quality). A comprehensive summary of the statistical results is provided in Table~\ref{tab:rq1}, and visual comparisons are shown in Figure~\ref{fig:quality1} and Figure~\ref{fig:quality2}.

\subsubsection{Issue Coverage}
First, regarding efficiency, participants identified significantly more issues when using CritiqueCrew ($M = 9.58, SD = 1.86$) compared to the Baseline tool ($M = 7.96, SD = 1.71$). Since the data violated normality assumptions, we conducted a Wilcoxon signed-rank test, which confirmed a significant improvement ($W = 49.0, p = .006, r = .59$).

Role Specialization Effect (Study 2). To isolate the source of this gain, Study 2 compared the Multi-Perspective mode against the Unified Expert mode. Results showed that the Multi-Perspective mode yielded significantly higher coverage ($M = 11.08, SD = 1.18$) than the Unified Expert mode ($M = 9.92, SD = 1.18$). This difference was statistically significant with a large effect size ($W = 9.0, p < .001, r = .82$). This suggests that decomposing critique into specialized roles encourages broader issue discovery.

\subsubsection{Solution Effectiveness}
The advantage of CritiqueCrew was even more pronounced in the critical metric of Solution Effectiveness. Blind scoring by three independent experts showed excellent inter-rater reliability (ICC(2,$k$) = 0.8), indicating highly consistent expert judgments.

Solutions produced with CritiqueCrew ($M = 5.22, SD = 0.98$) were rated significantly higher than those from the Baseline ($M = 4.30, SD = 0.47$). A paired $t$-test confirmed this improvement ($t(23) = 3.90, p < .001, d = 1.19$, 95\% CI [0.43, 1.40]).

Role Specialization Effect (Study 2). Study 2 further supported this finding: designs from the Multi-Perspective mode ($M = 5.77, SD = 0.49$) significantly outperformed those from the Unified Expert mode ($M = 5.02, SD = 0.67$). The difference was significant ($W = 5.0, p < .001, r = .85$), with similarly high inter-rater reliability (ICC(2,$k$) = 0.85). Experts noted that the multi-perspective feedback led to more strategic and holistic design optimizations.

\begin{table*}[t]
  \centering
  \small
  \renewcommand{\arraystretch}{1.2}
  \setlength{\tabcolsep}{6pt}
  \caption{Summary of RQ1 quantitative results on issue coverage and solution effectiveness.}
  \label{tab:rq1}

  \begin{tabular}{@{} l l p{0.42\textwidth} l l @{}}
    \toprule
    \textbf{Study} & \textbf{Metric} & \textbf{Descriptives} & \textbf{Test} & \textbf{Effect Size} \\
    \midrule

    1 & Issue Coverage &
    Base: $M=7.96, SD=1.71$ \quad CC: $M=9.58, SD=1.86$ &
    Wilcoxon $W=49.0$, $p=.006$ &
    $r=.59$ \\

    1 & Solution Effectiveness &
    Base: $M=4.30, SD=0.47$ \quad CC: $M=5.22, SD=0.98$ &
    $t(23)=3.90$, $p<.001$ &
    $d=1.19$ \\

    \midrule

    2 & Issue Coverage &
    Unified: $M=9.92, SD=1.18$ \quad Multi: $M=11.08, SD=1.18$ &
    Wilcoxon $W=9.0$, $p<.001$ &
    $r=.82$ \\

    2 & Solution Effectiveness &
    Unified: $M=5.02, SD=0.67$ \quad Multi: $M=5.77, SD=0.49$ &
    Wilcoxon $W=5.0$, $p<.001$ &
    $r=.85$ \\

    \bottomrule
  \end{tabular}

  \vspace{2pt}
  \footnotesize \textit{Note:} Base = Baseline (SpecAI), CC = CritiqueCrew, Unified = Unified Expert, Multi = Multi-Perspective.
\end{table*}

In summary, our results demonstrate that CritiqueCrew's multi-perspective paradigm broadly enhances both the quantity (Issue Coverage) and quality (Solution Effectiveness) of design outcomes. Crucially, Study 2 confirms that these gains are not merely artifacts of the interface, but are driven by the role orchestration mechanism itself. By decomposing critique into distinct functional lenses, the system helps designers move beyond surface-level fixes toward deeper, trade-off-aware optimization.

\subsection{RQ2: CritiqueCrew Improves Subjective User Experience}
\begin{figure*}[t]
\centering
\includegraphics[width=\textwidth]{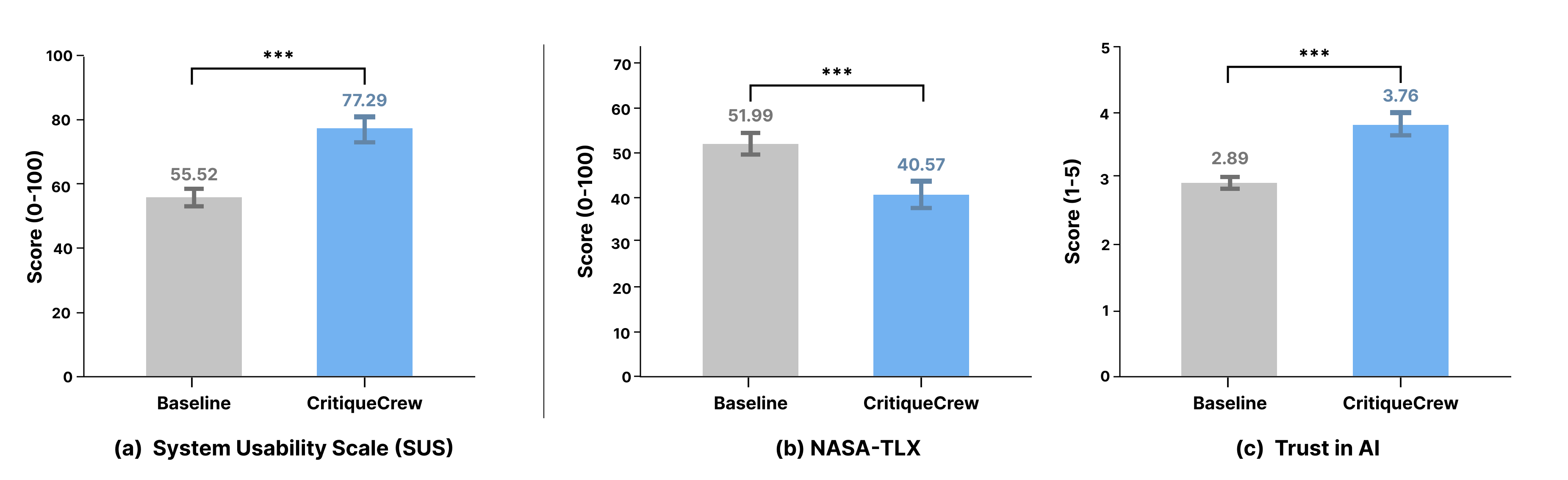} 
\caption{Comparison of subjective experience metrics between CritiqueCrew and the baseline tool (SpecAI). \textbf{(a) System Usability Scale (SUS, 0–100; Study 1 only). (b) NASA-TLX mental workload (0–100; lower is better). (c) Trust in AI (TAI, 1–5).} Panel (a) is visually separated from panels (b) and (c) because SUS was collected only in Study 1, whereas NASA-TLX and TAI ratings pool responses from all study sessions in which participants used the baseline tool and CritiqueCrew. Across all metrics, CritiqueCrew was perceived as more usable, imposed lower mental workload, and was trusted more than the baseline tool. ***$p$ < .001.}
\Description{Bar charts comparing subjective metrics. (a) SUS scores: CritiqueCrew (M=77.29) is significantly higher than Baseline (M=55.52). (b) NASA-TLX scores: CritiqueCrew (M=40.57) is significantly lower than Baseline (M=51.99). (c) Trust in AI scores: CritiqueCrew (M=3.76) is higher than Baseline (M=2.89). Error bars represent standard error. All differences are significant at p < .001.}
\label{fig:experience1}
\end{figure*}

\begin{figure*}[t]
\centering
\includegraphics[width=\textwidth]{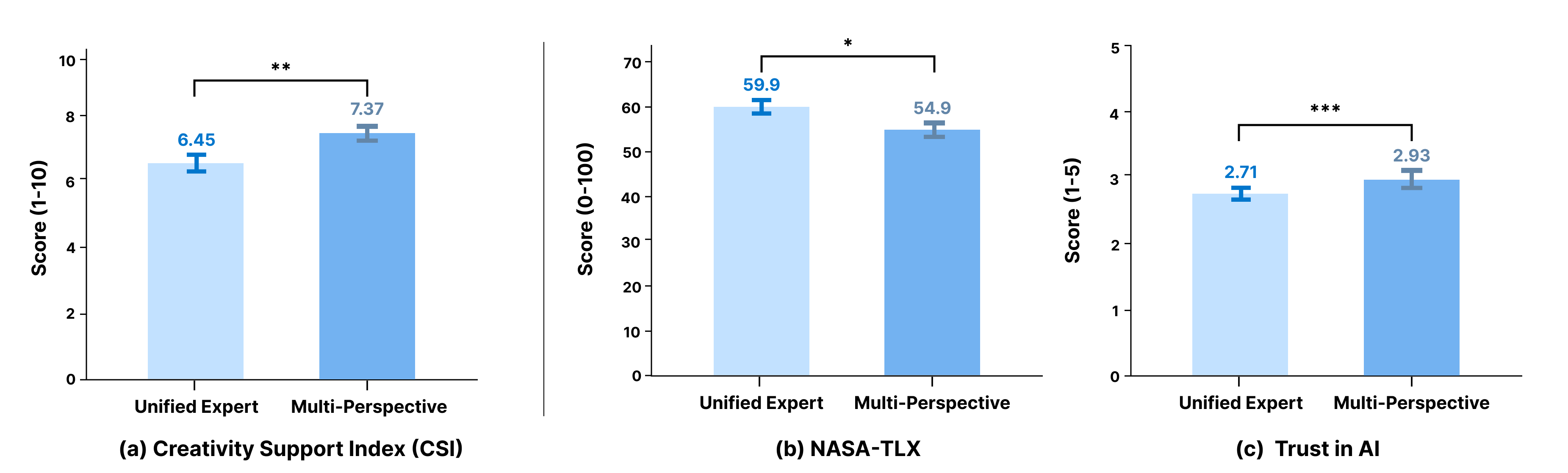} 
\caption{Comparison of subjective experience between the Unified Expert and Multi-Perspective modes of CritiqueCrew in Study 2. \textbf{(a) Creativity Support Index (CSI, 1–10; Study 2 only). (b) NASA-TLX mental workload (0–100; lower is better). (c) Trust in AI (TAI, 1–5).} Compared to the Unified Expert mode, the Multi-Perspective mode yielded higher CSI scores, slightly lower NASA-TLX workload, and higher Trust in AI ratings, indicating that participants experienced stronger creativity support, lower cognitive effort, and greater willingness to rely on the system’s suggestions when multiple expert roles were available. *$p$ < .05, **$p$ < .01, ***$p$ < .001.}
\Description{Bar charts comparing Unified Expert vs. Multi-Perspective modes. (a) CSI scores: Multi-Perspective (M=7.37) is higher than Unified Expert (M=6.45). (b) NASA-TLX scores: Multi-Perspective (M=54.9) is lower than Unified Expert (M=59.9). (c) Trust in AI scores: Multi-Perspective (M=2.93) is higher than Unified Expert (M=2.71). Significant differences are marked: * p < .05, ** p < .01, *** p < .001.}
\label{fig:experience2}
\end{figure*}

Our RQ2 examines the impact of CritiqueCrew's conversational workflow on designers' subjective experience. Questionnaire data showed that CritiqueCrew demonstrated significant advantages over the static baseline in system usability, cognitive load, and trust in AI (Study 1; see Figure~\ref{fig:experience1}), and that its multi-perspective configuration further optimized these metrics compared to the unified expert variant (Study 2; see Figure~\ref{fig:experience2}). Table~\ref{tab:rq2} summarizes the key statistics.

\begin{table*}[t]
  \centering
  \small
  \renewcommand{\arraystretch}{1.15}
  \setlength{\tabcolsep}{6pt}
  \caption{Summary of RQ2 quantitative results on subjective experience.}
  \label{tab:rq2}

  \begin{tabular}{@{} l l p{0.45\textwidth} l l l @{}}
    \toprule
    \textbf{Study} & \textbf{Metric} & \textbf{Descriptives} & \textbf{Test} & \textbf{$d$} & \textbf{95\% CI} \\
    \midrule

    1 & SUS &
    Base: $M=55.5, SD=9.9$;\; CC: $M=77.3, SD=15.2$ &
    $t(23)=6.37$, $p<.001$ &
    1.30 &
    $[15.1, 28.5]$ \\

    1 & NASA-TLX &
    Base: $M=52.0, SD=8.2$;\; CC: $M=40.6, SD=9.8$ &
    $t(23)=-5.24$, $p<.001$ &
    -1.07 &
    --- \\

    1 & TAI &
    Base: $M=2.89, SD=0.37$;\; CC: $M=3.76, SD=0.68$ &
    $t(23)=6.13$, $p<.001$ &
    1.25 &
    --- \\

    \midrule

    2 & NASA-TLX &
    Unified: $M=59.9, SD=6.2$;\; Multi: $M=54.9, SD=5.5$ &
    $t(23)=2.74$, $p=.012$ &
    0.56 &
    --- \\

    2 & TAI &
    Unified: $M=2.71, SD=0.28$;\; Multi: $M=2.93, SD=0.36$ &
    $t(23)=-3.81$, $p<.001$ &
    0.78 &
    --- \\

    2 & CSI &
    Unified: $M=6.45, SD=0.91$;\; Multi: $M=7.37, SD=0.69$ &
    $t(23)=-3.11$, $p=.005$ &
    0.64 &
    --- \\

    \bottomrule
  \end{tabular}

  \vspace{2pt}
  \footnotesize \textit{Note:} Base = Baseline (SpecAI), CC = CritiqueCrew, Unified = Unified Expert, Multi = Multi-Perspective. SUS = System Usability Scale; NASA-TLX = Task Load Index; TAI = Trust in AI; CSI = Creativity Support Index (as defined in the paper). Effect sizes are Cohen’s $d$; negative $d$ indicates lower scores for the second condition listed (CC or Multi), depending on the row.
\end{table*}

\subsubsection{System Usability Scale(SUS)}
In Study 1, CritiqueCrew achieved a mean SUS score of 77.29 ($SD = 15.20$), whereas the baseline tool received 55.52 ($SD = 9.92$). A paired $t$-test confirmed that this gap was highly significant ($t(23) = 6.37, p < .001, d = 1.30$). While the baseline score falls below the industry benchmark of 68, CritiqueCrew clearly exceeds this threshold, indicating a highly usable system with a lower learning curve.

\subsubsection{Cognitive Load (NASA-TLX)}
In Study 1, participants reported significantly lower cognitive load when using CritiqueCrew ($M = 40.57, SD = 9.75$) compared to the baseline ($M = 51.99, SD = 8.24$), achieving a \textbf{21.9\%} reduction ($t(23) = -5.24, p < .001, d = -1.07$).

\textbf{Role Specialization Effect (Study 2).} Study 2 further clarified the role of the architecture. The Multi-Perspective mode produced lower workload ($M = 54.9, SD = 5.47$) than the Unified Expert mode ($M = 59.9, SD = 6.24$). This difference was significant ($t(23) = 2.74, p = .012, d = 0.56$). These results suggest that despite introducing multiple viewpoints, the structured orchestration effectively manages complexity rather than increasing cognitive burden.

\subsubsection{Trust in AI (TAI)}
CritiqueCrew received significantly higher trust ratings ($M = 3.76, SD = 0.68$) than the baseline ($M = 2.89, SD = 0.37; t(23) = 6.13, p < .001, d = 1.25$).

\textbf{Role Specialization Effect (Study 2).} Multi-perspective critique yielded higher trust ($M = 2.93, SD = 0.36$) than the unified expert ($M = 2.71, SD = 0.28$). A paired $t$-test confirmed this effect ($t(23) = -3.81, p < .001, d = 0.78$), indicating that role-differentiated critiques foster greater user confidence.

\subsubsection{Creativity Support Index(CSI)}
Although CSI was not part of the baseline comparison, Study 2 revealed that multi-perspective critique significantly boosted creativity support ($M = 7.37, SD = 0.69$) compared to the unified expert ($M = 6.45, SD = 0.91$). A paired $t$-test confirmed the improvement ($t(23) = -3.11, p = .005, d = 0.64$).

CritiqueCrew significantly outperforms the static baseline across all usability and trust metrics. Crucially, Study 2 reveals that these experiential gains are amplified by the multi-perspective architecture. Rather than increasing complexity, the structured orchestration of expert roles reduces cognitive load and enhances creativity support, suggesting that designers prefer the clarity of role-differentiated guidance over a generic, unified response.

\subsection{RQ3: CritiqueCrew Fosters Strategic Design Partnership}

Our qualitative analysis reveals how designers perceived CritiqueCrew’s multi-perspective critique and how this conversational paradigm fundamentally changed their design optimization process. We identified three core themes.

\subsubsection{From Surface Fixing to Cross-Functional Strategy}
CritiqueCrew catalyzed a cognitive shift among designers, transforming their process from executing surface-level corrections to engaging in strategic thinking. Unlike the static baseline which engendered a linear, ``checklist-style'' workflow, CritiqueCrew fostered active decision-making by presenting constructive conflicts. For instance, when the Product Vision role suggested a prominent brand-colored button while the UX role flagged its insufficient contrast, designers were forced to weigh trade-offs rather than passively fixing isolated issues. As P4 (Study 1) noted, this prompted him to re-evaluate the entire design: \textit{``CritiqueCrew helped me identify that the `eMall' page was missing crucial product price information, which prompted me to re-evaluate the entire design from the perspectives of information architecture and business logic integrity.''}

Study 2 reinforced the value of this multi-perspective approach. Participants explicitly noted that divergent viewpoints felt ``more convincing and realistic'' than the Unified Expert. As P1 (Study 2) explained, the unified critic provided \textit{``cleaner but generic advice,''} whereas the multi-perspective system acted as a \textit{``comprehensive safety net.''} However, this depth came with a trade-off. Execution-oriented designers like P17 (Study 2) noted that conflicting views could sometimes increase cognitive load for simple tasks: \textit{``Sometimes the multi-perspective information is too much... I don't care what the engineer thinks at this stage; I just want to finish the design.''} This suggests that while multi-faceted perspectives are superior for strategic quality, they require more active processing.

Ultimately, this workflow crystallizes the AI's transformed role. As P3 (Study 1) described: \textit{``I tend to first let it evaluate the interface, and then I modify it by combining its feedback from various dimensions with my own design experience.''} This recasts the AI from a prescriptive auditor to a strategic partner.

\subsubsection{Contextual Remediation Preserves Creative Flow}

A critical advantage of CritiqueCrew was its ability to integrate feedback directly onto the design canvas, addressing the fragmented ``read--understand--locate--modify'' loop of the baseline. Participants universally praised the practical benefits. P2 (Study 1) highlighted the \textit{``one-click copy text feature''} for eliminating tedious parsing, while P6 (Study 1) noted that while the baseline required him to \textit{``manually select and locate''} components, CritiqueCrew automated this through its \textit{``one-click preview.''}

By minimizing the cognitive cost of execution, the system allowed designers to remain in a state of creative flow. P2 (Study 1) explained that with other tools, she would prefer to \textit{``manually revise it once and then let the tool help me evaluate the revised draft''} to protect her concentration. CritiqueCrew’s seamless integration makes such defensive behaviors unnecessary, transforming real-time feedback from a potential disturbance into an efficient accelerator.

\subsubsection{Structured Dialogue Builds Trust and Agency}

Trust in CritiqueCrew was driven by explainability and agency. Unlike the baseline's opaque suggestions, CritiqueCrew's role-based rationale helped designers understand the \textit{``why''} behind changes. P7 (Study 1) noted that the baseline's vague feedback often forced him to waste time validating issues: \textit{``I had to spend time thinking about whether there was really a problem.''}

Study 2 revealed that role differentiation was key to this trust. Comparing the multi-perspective system to the unified expert, participants felt that distinct personas helped them form clearer mental models of the AI's reasoning. Log analysis of Study 2 showed that designers actively used the \texttt{@} feature to query specific roles (52\% to UX Designer, 32\% to Product Manager, 17\% to Engineer), treating them as specialized teammates. When roles disagreed, designers engaged in negotiation to find a ``middle ground'' or used the conflict as creative provocation. Some designers, like P8 (Study 2), would reject conflicting suggestions outright (\textit{``There are some suggestions I won’t refer to''}), yet still found value in the debate: \textit{``It gave me a new idea.''} This confirms that genuine trust is forged not through AI infallibility, but by empowering designers to confidentially direct the collaborative process.

\section{Discussion}

This study evaluated CritiqueCrew, a conversational tool designed to transform AI from a passive auditor into a strategic critique partner. Our results demonstrate that CritiqueCrew's multi-perspective orchestration significantly improves both design quality (RQ1) and user experience (RQ2) compared to static baselines. Crucially, Study 2 confirms that these benefits stem from the structured interplay of expert roles, rather than just underlying model capabilities. In this section, we interpret these findings through the lens of design theory, compare our approach with existing methods, and outline future directions.

\subsection{Principal Findings: From Validation to Reflective Dialogue}
Our findings indicate that CritiqueCrew moves beyond the traditional ``validation'' paradigm of static tools towards a ``reflective dialogue.'' First, the system improved design quality not just by fixing more issues, but by fostering deeper strategic thinking. Study 2 showed that decomposing critique into specialized roles (UX, PM, Engineer) yielded higher issue coverage and solution effectiveness than a unified expert model. This suggests that the productive tension \cite{bucciarelli1988ethnographic} between conflicting roles helps designers uncover blind spots that a single ``correct'' answer might obscure.

Second, despite the increased information density, CritiqueCrew reduced cognitive load and increased trust. Participants did not perceive multi-perspective feedback as ``noise,'' but as a structured framework for reasoning about trade-offs. This aligns with Distributed Cognition theory \cite{hollan2000distributed}, where intelligence emerges from the negotiation between perspectives. By externalizing this negotiation, CritiqueCrew makes the ``black box'' of design trade-offs visible and manipulable, helping designers move from surface-level fixes to holistic optimization.

\subsection{Implications for Design}
Our study offers three key implications for future AI-empowered creative tools:
\begin{enumerate}

    \item \textbf{Beyond problem detection, empower solution exploration.} Future tools should shift focus from \textit{``what’s wrong''} to \textit{``how it can be improved.''} Our findings suggest that AI can act as a catalyst for creativity, aligning with UIDEC~\cite{shokrizadeh2025dancing} where constraints serve as creative scaffolds. CritiqueCrew achieves this by translating abstract rules into the \textit{``stances''} of different expert roles. The higher Creativity Support scores in Study 2 confirm that this multi-perspective approach not only aids in identifying issues but also supports the generation of strategic, trade-off-aware solutions, empowering designers to move beyond mere compliance.
    
    \item \textbf{Leverage Role-Based Interaction Paradigms.} Using a multi-perspective system broadens critique dimensions and serves as an effective interaction paradigm. Converting abstract rules (e.g., WCAG) into recommendations from concrete roles (e.g., ``Accessibility Expert'') better aligns with designers’ mental models. However, this personification introduces risks of over-trust. Our findings suggest a mitigation: \textbf{deliberately surfacing constructive conflicts}. By presenting differing viewpoints, such as those from the Product Vision and User Experience roles, the system compels the designer to act as the final arbiter. In our study, designers actively used the @-mention feature to question and cross-check recommendations, illustrating how role differentiation and conversational control help keep human judgment at the center rather than being displaced by automation.
    
    \item \textbf{Deep contextual integration is key to trust.} 
    Tools must seamlessly integrate into existing workflows. Our study indicates that direct on-canvas highlighting and real-time previews minimize context switching, allowing designers to view AI as a natural extension of their process. However, this reliance on LLMs is a double-edged sword. The efficiency of ``one-click apply'' features can amplify the negative impact of hallucinations. CritiqueCrew addresses this through explainability and conversational clarification. By allowing designers to probe reasoning (e.g., \textit{``@Engineer}, why is this costly?''), we empower them to scrutinize the AI's logic. This suggests that future tools must be designed for scrutability, tying explanations and actions directly back to the shared artifact to minimize the risks of blind adoption.
\end{enumerate}

\subsection{Limitations and Future Work}
While our results are promising, we acknowledge specific limitations in our study design and system scope. 

First, although Study 2 was designed to primarily examine the contribution of role-based orchestration, the Multi-Perspective condition also externalizes synthesis through an explicit Lead Coordinator. Our unified expert baseline similarly aggregates and prioritizes feedback, but this integration occurs implicitly within a single model response. Therefore, the observed improvements likely reflect the combined effects of role-specialized reasoning and explicit coordination, rather than any single mechanism alone. Future work could further disentangle these factors via ablation studies or factorial designs that independently vary role specialization and coordination structure.

Second, our evaluation relied on pre-defined mockups rather than designers' own creative work. While this ensured rigorous experimental control and comparability across conditions, it may not fully capture the emotional attachment and deep context designers have with their own creations. In real-world settings, designers might be more defensive or nuanced in accepting critique. However, participants noted that the mockups were realistic, and the task constraints effectively simulated common industry scenarios.

Third, our evaluation focused on individual designers. While designers often serve as the primary bridge between disciplines, we did not directly observe how real product managers or engineers would interact with the system. This focus was intentional to assess the tool's value for its primary user group, but future work should broaden the scope to include diverse stakeholders to validate the fidelity of our expert personas.

Fourth, although our sample size ($N=48$ across two studies) provided sufficient statistical power for detecting large effects, broader generalization would benefit from a larger and more diverse participant pool. Additionally, our operationalization of design quality relied on expert ratings of final static mockups. While this is a standard practice in HCI, it does not capture downstream effects such as end-user behavior conversion rates or long-term code maintainability, which would require longitudinal deployment to assess.

Finally, CritiqueCrew currently evaluates single frames. The current implementation instantiates a fixed set of three roles and does not yet adapt dynamically to different task types. Future iterations will expand to support user flow analysis across multiple screens.

Future work will proceed in three strategic directions. First, we plan to conduct a \textit{longitudinal field study} to deploy CritiqueCrew within real design teams. This will allow us to examine how conflict frequency and negotiation strategies evolve over long-term projects, and how AI-generated critiques influence design ownership over time. Second, we aim to expand the system's contextual awareness beyond single frames to support user flow analysis. By enabling the system to understand transitions and interaction logic across multiple screens, we can provide more holistic feedback on information architecture and journey consistency. Finally, we envision a customizable expert panel where designers can configure the personas or even create custom roles tailored to their specific project needs. This would transform CritiqueCrew from a fixed tool into an adaptable, evolving design partner.

\section{Conclusion}
We introduced CritiqueCrew to reimagine AI's role in design critique—shifting from a static ``problem auditor'' to a dynamic ``solution co-creator.'' By orchestrating multi-perspective expert roles within a structured workflow, the system bridges the gap between abstract feedback and creative action. CritiqueCrew embeds critique directly into the design process, transforming isolated problem lists into actionable, explainable, and context-aware solutions.

Across two controlled studies ($N=48$), CritiqueCrew significantly outperformed both a state-of-the-art static tool and a unified expert baseline. Designers using our system resolved more issues, produced higher-quality outcomes, and reported significantly lower cognitive load and higher trust. Crucially, our findings validate that these gains stem from the role orchestration mechanism itself: by surfacing productive tension between distinct perspectives (e.g., UX vs. Engineering), the system catalyzes deeper strategic thinking rather than mere error correction.

This work contributes a novel paradigm for human--AI collaboration in design: moving beyond automated checking toward orchestrated, multi-perspective dialogue. It offers actionable insights for the next generation of creative support tools, demonstrating that structured role differentiation is key to building trustworthy and effective AI partners in complex, cross-functional domains.

\begin{acks}
This research was supported by the Foundation of National Key Laboratory of Human Factors Engineering (No. HFNKL2024W03).
\end{acks}

\bibliographystyle{ACM-Reference-Format}
\bibliography{software}

\clearpage
\appendix

\section{Formative Interview Study}
The formative study participants' detailed information (gender, age, role, and years of design
experience) is shown in Table~\ref{tab:participants}. All participants had prior experience 
using Figma in their daily work and had participated in design critiques as well as the full 
workflow from product design to implementation.

\begin{table}[t]
\centering
\small
\setlength{\tabcolsep}{5pt}
\renewcommand{\arraystretch}{1.15}
\caption{Participant demographics and professional roles.}
\label{tab:participants}

\begin{tabular}{@{} c c c c p{0.45\columnwidth} @{}}
\toprule
\textbf{PID} & \textbf{Gender} & \textbf{Age} & \textbf{Experience} & \textbf{Professional Role} \\
\midrule
P1  & Female & 18--24 & 2 years & UI Designer \\
P2  & Female & 32--38 & 7 years & UI Designer \\
P3  & Female & 32--38 & 7 years & UI Designer \\
P4  & Female & 25--31 & 2 years & Interaction Designer \\
P5  & Female & 25--31 & 3 years & Interaction Designer \\
P6  & Female & 25--31 & 5 years & Interaction Designer \\
P7  & Male   & 25--31 & 2 years & Product Manager \\
P8  & Female & 25--31 & 4 years & Product Manager \\
P9  & Male   & 25--31 & 4 years & Product Manager \\
P10 & Male   & 25--31 & 3 years & Software Engineer (Front-end) \\
P11 & Female & 25--31 & 5 years & Software Engineer (Front-end) \\
P12 & Male   & 25--31 & 5 years & Software Engineer (Front-end) \\
\bottomrule
\end{tabular}
\end{table}

\section{System Prompts}
\label{sec:prompts}

This section presents the detailed system prompts used for each expert module and the lead coordinator in CritiqueCrew.

\begin{framed}
\noindent\textbf{User Experience Module Prompt}

\medskip
\noindent\textbf{Role \& Objective}\\
As a User Experience Expert (UX researcher / UI designer / accessibility specialist), evaluate \emph{usability}, \emph{accessibility}, and \emph{overall user experience}, and provide specific, actionable recommendations.

\medskip
\noindent\textbf{Inputs}\\
1) UI screenshot\\
2) UI JSON\\
3) Product context (goals, brand keywords, theme colors, target users)

\medskip
\noindent\textbf{Output Format (JSON)}\\
\begin{lstlisting}
{
  "feedback_id": "unique_id",
  "source_role": "user_experience",
  "issues": [
    {
      "nodeId": "element_id",
      "nodeName": "element_name",
      "elementType": "TEXT|FRAME|VECTOR|...",
      "issueType": "usability|accessibility|navigation|...",
      "severity": "critical|high|medium|low",
      "description": "Issue description",
      "remediation": {
        "action": "Action recommendation",
        "specificSuggestion": "Specific improvement",
        "technicalSolution": "Implementation plan"
      }
    }
  ],
  "summary": {...},
  "priority": "critical|high|medium|low",
  "detailed_analysis": "Overall analysis"
}
\end{lstlisting}

\medskip
\noindent\textbf{Analysis Scope}\\
1) \emph{Usability}: information clarity, task flow, interaction efficiency, consistency. \\
2) \emph{Accessibility} (WCAG 2.1 AA \& AAA): text contrast ratio ($\geq$4.5:1 for normal, $\geq$3:1 for large), minimum touch target size (44×44px), readable font size ($\geq$16px), keyboard navigation, semantic labels. \\
3) \emph{Mobile Optimization}: responsive design, touch/gesture experience, loading performance. \\
4) \emph{\ldots}

\medskip
\noindent\textbf{Execution Rules}\\
- Focus only on interface and interaction issues. \\
- Each issue must include \texttt{nodeId} and describe user impact. \\
- When data is missing, provide conservative, general suggestions and ensure valid JSON output.
\end{framed}

\begin{framed}
\noindent\textbf{Product Vision Module Prompt}

\medskip
\noindent\textbf{Role \& Objective}\\
As a Product Vision Expert (product manager / business strategist), evaluate whether the design aligns with \emph{business goals}, \emph{brand identity}, and \emph{market positioning}, and provide actionable recommendations.

\medskip
\noindent\textbf{Inputs}\\
1) UI screenshot \\
2) UI JSON \\
3) Product context (goals, brand keywords, theme colors, target users)

\medskip
\noindent\textbf{Output Format (JSON)}\\
\begin{lstlisting}
{
  "feedback_id": "unique_id",
  "source_role": "product_vision",
  "issues": [
    {
      "nodeId": "element_id",
      "nodeName": "element_name",
      "elementType": "TEXT|FRAME|VECTOR|...",
      "issueType": "business_goal|brand|content|market|...",
      "severity": "critical|high|medium|low",
      "description": "Issue description",
      "remediation": {
        "action": "Action recommendation",
        "specificSuggestion": "Specific improvement",
        "technicalSolution": "Implementation plan"
      }
    }
  ],
  "summary": {...},
  "priority": "critical|high|medium|low",
  "detailed_analysis": "Overall analysis"
}
\end{lstlisting}

\medskip
\noindent\textbf{Analysis Scope}\\
1) \emph{Business Goal Alignment}: support of product goals, conversion flow optimization, prioritization of key functions, potential impact on KPIs. \\
2) \emph{Brand Consistency}: correct use of theme colors, alignment with brand keywords (e.g., professional, simple, trustworthy), visual hierarchy, emotional expression of brand values. \\
3) \emph{Content Strategy}: clarity and readability of copy, effectiveness of CTAs, information prioritization, guidance for user flows. \\
4) \emph{Market Positioning}: differentiation, alignment with user expectations in target market, competitive advantage. \\
5) \emph{\ldots}

\medskip
\noindent\textbf{Execution Rules}\\
- Each issue must include \texttt{nodeId} and describe its business/brand impact. \\
- Provide recommendations grounded in product strategy and brand identity. \\
- Focus on how design choices affect user conversion and brand communication. \\
- When data is missing, give conservative, general suggestions and ensure valid JSON output.
\end{framed}


\begin{framed}
\noindent\textbf{Engineering Module Prompt}

\medskip
\noindent\textbf{Role \& Objective}\\
As an Engineering Expert (front-end / software engineer perspective), evaluate the \emph{engineering feasibility}, \emph{implementation cost}, and \emph{performance implications} of the design. Provide concrete, actionable technical solutions.

\medskip
\noindent\textbf{Inputs}\\
1) UI screenshot \\
2) UI JSON \\
3) Product context (goals, brand keywords, theme colors, target users)

\medskip
\noindent\textbf{Output Format (JSON)}\\
\begin{lstlisting}
{
  "feedback_id": "unique_id",
  "source_role": "engineering",
  "issues": [
    {
      "nodeId": "element_id",
      "nodeName": "element_name",
      "elementType": "TEXT|FRAME|VECTOR|...",
      "issueType": "feasibility|performance|accessibility|...",
      "severity": "critical|high|medium|low",
      "description": "Issue description",
      "remediation": {
        "action": "Action recommendation",
        "specificSuggestion": "Specific improvement",
        "technicalSolution": "Implementation plan / code-level suggestion"
      }
    }
  ],
  "summary": {...},
  "priority": "critical|high|medium|low",
  "detailed_analysis": "Overall technical analysis"
}
\end{lstlisting}

\medskip
\noindent\textbf{Analysis Scope}\\
1) \emph{Engineering Feasibility}: CSS/JS complexity, third-party dependencies, browser/device compatibility, responsive implementation challenges. \\
2) \emph{Performance Impact}: image optimization, font loading, CSS/JS complexity, rendering cost (FCP, LCP, CLS, FID). \\
3) \emph{Accessibility Implementation Support}: technical fixes for issues flagged by the UX Module (contrast, touch target size, keyboard navigation, screen reader compatibility). \\
4) \emph{Development Cost Estimation}: effort by complexity level (minor style tweak vs. new component/architecture-level change). \\
5) \emph{Design Breakdown Detection}: identify implementation inconsistency, over-complex interactions, or missing accessibility support. \\
6) \emph{\ldots}

\medskip
\noindent\textbf{Execution Rules}\\
- Focus on feasibility, cost, and performance rather than pure usability. \\
- Each issue must include \texttt{nodeId}, technical challenges, cost estimate, and alternative solutions when applicable. \\
- Provide realistic implementation details (CSS/JS examples where useful). \\
- Collaborate with the UX Module by supplying concrete fixes for accessibility issues.
\end{framed}


\begin{framed}
\noindent\textbf{Lead Coordinator Prompt}

\medskip
\noindent\textbf{Role \& Objective}\\
As the Lead Coordinator, you act as the \emph{facilitator} in the Conversational Design Critique.  Your goal is to empower designers to make informed decisions, not to decide for them.  You synthesize inputs from three expert modules (User Experience, Product Vision, Engineering).

\medskip
\noindent\textbf{Inputs}\\
1) Feedback and issues identified by the three expert modules \\
2) Product context and goals \\
3) Structured critique agenda items from modules

\medskip
\noindent\textbf{Output Format (JSON)}\\
\begin{lstlisting}
{
  "conversational_opening": "Lead Coordinator introduction",
  "overall_score": 1-10,
  "agenda_items": [
    {
      "priority": "critical|high|medium|low",
      "title": "Agenda title",
      "component_group": "UI component(s)",
      "affected_roles": ["user_experience", "product_vision", "engineering"],
      "issue_summary": "Summary of the issue",
      "recommendation": "Recommended actions",
      "conflicts": [
        {
          "conflicting_roles": ["module1", "module2"],
          "conflict_description": "Conflict description",
          "tradeoff_question": "Open trade-off question"
        }
      ],
      "estimated_effort": "Workload estimate"
    }
  ],
  "conflicts_to_surface": [...],
  "component_analysis": [...],
  "positive_highlights": ["..."],
  "next_conversation_points": ["..."]
}
\end{lstlisting}

\medskip
\noindent\textbf{Analysis Scope}\\
1) \emph{Synthesize \& Thematize}: group feedback by components, highlight cross-cutting themes, avoid simple listing. \\
2) \emph{Prioritize}: order issues by severity (critical accessibility > core flow blockers > business misalignment > technical debt > aesthetic refinements). \\
3) \emph{Surface Conflicts}: when modules disagree, present trade-offs and guide designers with open-ended questions.

\medskip
\noindent\textbf{Execution Rules}\\
- Act as a facilitator, not the decision-maker. \\
- Use natural, conversational, and encouraging language. \\
- Provide a structured agenda to avoid cognitive overload. \\
- Highlight conflicts explicitly, phrased as questions (e.g., “How might we balance accessibility standards with brand consistency here?”). \\
- Emphasize collective perspectives (“we observed…”, “the team suggests…”) rather than absolute judgments.

\end{framed}

\section{Application Cases}

\begin{figure*}[t]
\centering
\includegraphics[width=\textwidth]{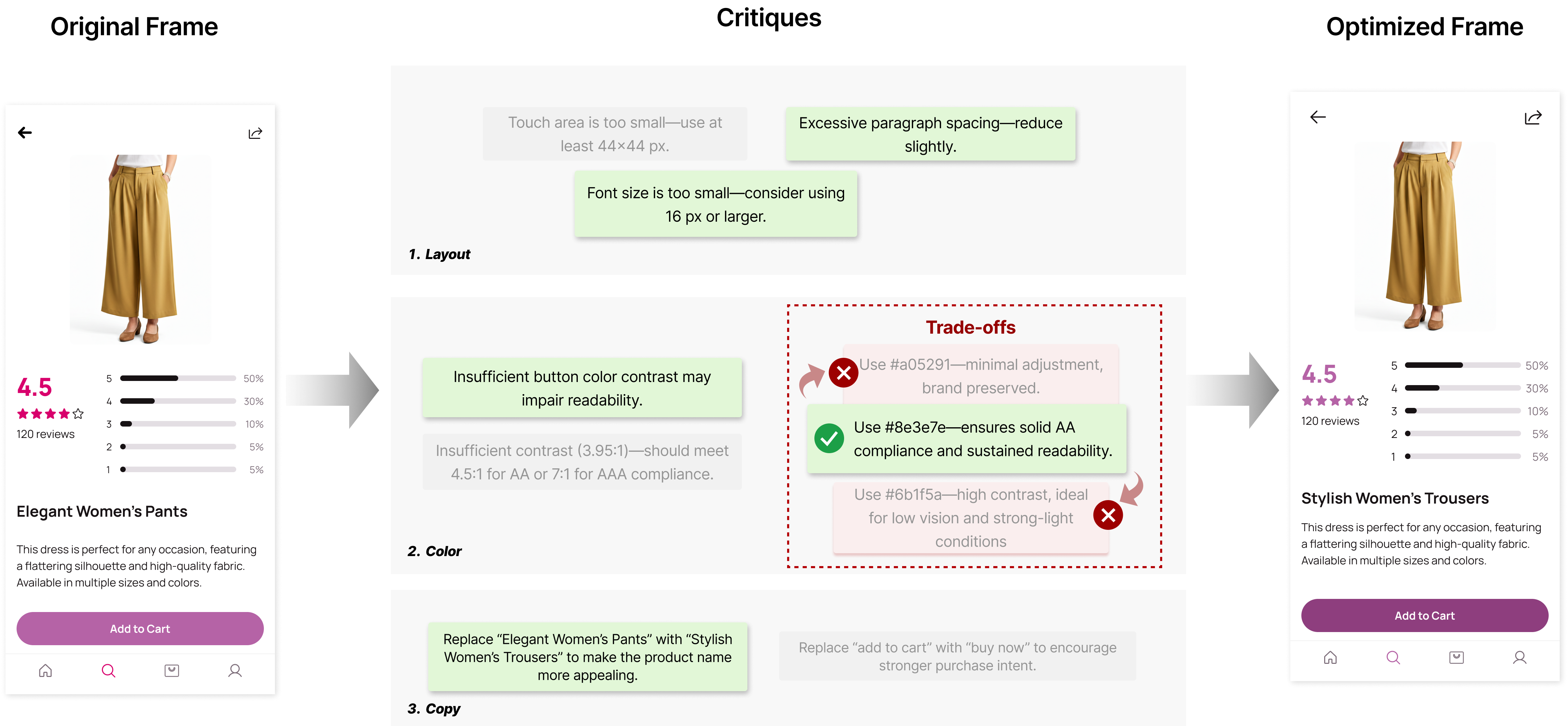} 
\caption{A case study of CritiqueCrew applied to a single frame of the e-commerce product page mockup. The system provides critiques across three dimensions: Layout, Color, and Copy. Notably, the color contrast suggestions are presented as ``Trade-offs,'' empowering the designer to make an informed decision between brand consistency and accessibility compliance, leading to the optimized frame on the right.}
\Description{The figure shows a three-step transformation of a mobile e-commerce product page. The left panel displays the original design with low-contrast buttons, small text, and the product titled “Elegant Women’s Pants.” The middle panel shows CritiqueCrew’s feedback across layout, color, and copy, including a highlighted “Trade-offs” box suggesting accessible color alternatives. The right panel presents the optimized frame with improved contrast, clearer spacing, larger text, and updated copy reading “Stylish Women’s Trousers.” Arrows indicate progression from original, to critiques, to optimized design.}
\label{fig:case1}
\end{figure*}

\begin{figure*}[t]
\centering
\includegraphics[width=\textwidth]{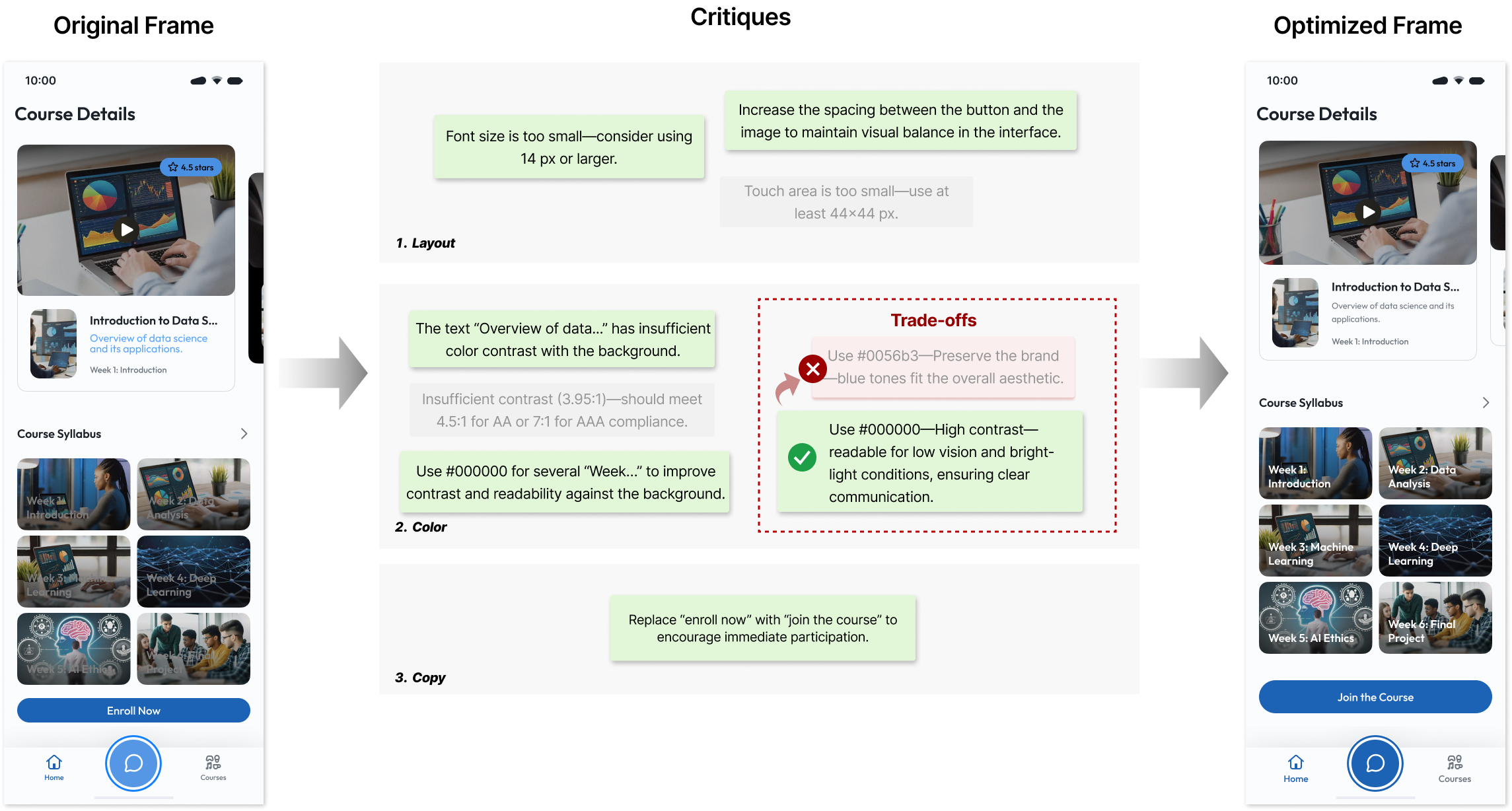} 
\caption{A case study of CritiqueCrew applied to a single frame from the online course details page mockup. The system identifies design issues from multiple dimensions. The color critique for text contrast is presented as ``Trade-offs,'' clearly illustrating the conflict between brand aesthetics and readability to guide the designer toward the final optimized version.}
\Description{The figure shows a three-step transformation of a mobile online-course details page. The left panel presents the original design with small text, low-contrast labels, and a blue “Enroll Now” button. The middle panel summarizes CritiqueCrew’s layout, color, and copy feedback, including a “Trade-offs” box comparing a brand-blue text color with a higher-contrast black alternative. The right panel shows the optimized frame with larger text, improved spacing, clearer contrast, and revised wording such as “Join the Course.” Arrows illustrate the progression from the original frame to critiques and then to the optimized design.}
\label{fig:case2}
\end{figure*}

To further illustrate the system's workflow and value, this section presents detailed application cases drawn from the study's e-commerce (Figure~\ref{fig:case1}) and online education (Figure~\ref{fig:case2}) scenarios. Each example focuses on the optimization of a single, representative frame, demonstrating how CritiqueCrew transforms abstract multi-perspective evaluations into concrete, actionable design modifications and thereby assists designers in efficiently improving design quality.

\section{Qualitative Analysis Codebook}
\label{sec:appendix_codebook}

Table \ref{tab:codebook} details the codebook developed during our formative study. It outlines the three core themes, their associated sub-codes, definitions, and representative quotes from participants.

\begin{table*}[t!]
\centering
\small 
\renewcommand{\arraystretch}{1.1}
\caption{The thematic analysis codebook derived from the formative study interviews (N=15).}
\label{tab:codebook}
\begin{tabular}{@{} p{2.5cm} p{3.2cm} p{0.30\textwidth} p{0.34\textwidth} @{}}
\toprule
\textbf{Theme} & \textbf{Sub-Code} & \textbf{Definition} & \textbf{Representative Quote} \\
\midrule

\multirow{15}{=}{\textbf{Negotiation of Competing Constraints}} 
& \textbf{Technical Debt vs. Visual Novelty} 
& Tensions where unique visual designs require non-standard engineering implementation, raising maintenance costs. 
& \textit{``The designer wanted a `Semibold' font for a popup title... I told them we don't have this font package. Supporting it means extra dev cost. It's not impossible, but I refused to avoid opening a floodgate for non-standard requests.''} (P10, Front-end Engineer) \\
\cmidrule(l){2-4}

& \textbf{Business KPIs vs. Design Consistency} 
& Conflicts where aggressive business goals (e.g., traffic entry) force violations of established UI component guidelines. 
& \textit{``The PM insisted on placing the `Robot' feature at the top layer for visibility... But our platform guidelines don't support this component. We argued for a long time and finally compromised by folding it into a menu.''} (P2, UI Designer) \\
\cmidrule(l){2-4}

& \textbf{Logic Gaps in Edge Cases} 
& The reliance on downstream roles (Dev/AI) to identify logical holes in product requirements that PMs miss. 
& \textit{``As a PM, I might miss edge cases, like a billing date falling on a non-work day at month-end. AI could help exhaust these scenarios that are hard for humans to fully cover.''} (P7, Product Manager) \\
\midrule

\multirow{15}{=}{\textbf{Trust \& Explainability}} 
& \textbf{Evidence-Based Rationale} 
& Users trust AI feedback only when it is grounded in cited sources (e.g., guidelines, psychology theories) rather than opaque opinions. 
& \textit{``What builds my trust is when the AI provides a basis for its suggestions... like citing a specific psychology theory. If it has a reference, I trust it; otherwise, I suspect hallucinations.''} (P5, Interaction Designer) \\
\cmidrule(l){2-4}

& \textbf{Contextual Grounding (PRD)} 
& The requirement for AI to ingest project-specific documents (PRD) to avoid generic, unhelpful advice. 
& \textit{``If the AI doesn't know our internal `black slang' or specific business context... it will just generate 99\% correct but useless nonsense. It must understand the PRD.''} (P8, Product Manager) \\
\cmidrule(l){2-4}

& \textbf{White-Box Reasoning} 
& The need for AI to reveal its decision-making process, especially when resolving conflicts. 
& \textit{``It shouldn't just give a final result. It must show the reasoning process... If it acts as a black box, I won't trust its judgment on conflicting opinions.''} (P6, Interaction Designer) \\
\midrule

\multirow{10}{=}{\textbf{Workflow Integration}} 
& \textbf{Phase-Dependent Utility} 
& The distinct shift in needs: ``Open-ended inspiration'' in early stages vs. ``Strict compliance checks'' in late stages. 
& \textit{``In the early concept phase, I want open-ended suggestions for inspiration. But in the late delivery phase, I trust human constraints more... AI is best for checking details like color contrast compliance.''} (P5, Interaction Designer) \\
\cmidrule(l){2-4}

& \textbf{Empowerment \& Explanation} 
& The need for AI to explain the ``Why'' behind a conflict and offer trade-offs, leaving the final decision to the human. 
& \textit{``The AI shouldn't just give a final result. It needs to show the reasoning process... acting like a moderator that explains the trade-offs between different perspectives.''} (P3, UI Designer) \\
\bottomrule
\end{tabular}
\end{table*}

\end{document}